%% file: main.tex
\documentclass[acmsmall]{acmart}
\usepackage{tikz}
\usetikzlibrary{positioning}
\usepackage{array} 
\usepackage{booktabs}
\usepackage{multirow}
\usepackage{lscape}
\usepackage{graphicx}
\usepackage{forest}
\usepackage{smartdiagram}
\usepackage{amsmath}
\usepackage{mathtools}
\usepackage[utf8]{inputenc}
\usepackage{dirtytalk}
\usepackage{array}
\usepackage{tabularx}
\usepackage{subfigure}

\usepackage{rotating}
\usepackage{float}
\usepackage{array, multicol, multirow, makecell}

\usepackage{bigstrut}
\newcolumntype{b}{X}
\newcolumntype{s}{>{\hsize=.4\hsize}X}

\AtBeginDocument{%
  \providecommand\BibTeX{{%
    \normalfont B\kern-0.5em{\scshape i\kern-0.25em b}\kern-0.8em\TeX}}}
    
\makeatletter
\newlength\mylena
\newlength\mylenb
\newcommand\mystrut[1][2]{%
    \setlength\mylena{#1\ht\@arstrutbox}%
    \setlength\mylenb{#1\dp\@arstrutbox}%
    \rule[\mylenb]{0pt}{\mylena}}
\makeatother    

\newcommand{\citePS}[1]{\cite{#1}}

\newcommand{\flavia}[1]{\textcolor{black}{#1}}
\newcommand{\vero}[1]{\textcolor{black}{#1}}
\newcommand{\adri}[1]{\textcolor{black}{#1}}

\begin{document}

\title{Use of Context in Data Quality Management: \\
a Systematic Literature Review}

\author{Flavia Serra} %\authornote{}
\email{fserra@fing.edu.uy}
\orcid{0000-0001-8588-798X}
\affiliation{
  \institution{Universidad de la República, Montevideo, Uruguay.}
  \institution{Université de Tours, Blois, France}
}

\author{Veronika Peralta}
\email{veronika.peralta@univ-tours.fr}
\orcid{0000-0002-9236-9088}
\affiliation{
  \institution{Université de Tours, Blois, France}
}

\author{Adriana Marotta}
\email{amarotta@fing.edu.uy}
\orcid{0000-0001-6547-466X}
\affiliation{
  \institution{Universidad de la República, Montevideo, Uruguay}
}

\author{Patrick Marcel}
\email{Patrick.Marcel@univ-tours.fr}
\orcid{0000-0003-3171-1174}
\affiliation{
  \institution{Université de Tours, Blois, France}
}

\renewcommand{\shortauthors}{F. Serra et al.}

\begin{abstract}
   The importance of context in data quality (DQ) was shown many years ago and nowadays is widely accepted. Early approaches and surveys defined DQ as \textit{fitness for use} and showed the influence of context on DQ. This paper presents a Systematic Literature Review (SLR) for investigating how context is taken into account in recent proposals for DQ management. We specifically present the planning and execution of the SLR, the analysis criteria and our results reflecting the relationship between context and DQ in the state of the art and, particularly, how that context is defined and used for DQ management.
\end{abstract}



\maketitle

\input{_intro.tex}

\input{_background.tex}

\input{_slr.tex}

\input{_results.tex}
\input{_examples.tex}

\input{_answers.tex}

\input{_conclusions.tex}

\bibliographystyle{ACM-Reference-Format}
\bibliography{references,references_SLR}

%\bibliographystylePS{ACM-Reference-Format}
%\bibliographyPS{references_SLR}

\end{document}

%% file: _intro.tex
\section{Introduction}
\label{sec:introduction}
Data Quality (DQ) is a very wide research area, which involves many different aspects, problems and challenges. The growing need to discover and integrate reliable information from heterogeneous data sources, distributed in the Web, Social Networks, Cloud platforms or Data Lakes, makes DQ an unavoidable topic, particularly hot in recent years (see e.g., \cite{chu_data_2016, sadiq_data_2018, pena_discovery_2019, bertossi_database_2019}). In addition, DQ has enormous relevance for the industry, due to its great impact on information systems in all application domains. 

\flavia{%As an example, 
Advances in information and communication technologies have led organizations %of digital government 
to manage increasingly large amounts of data. These generates interesting opportunities, both for the daily operations %of government organizations 
as well as for decision-making and strategic planning. 
%This is the reality of the e-Government Agency and Information and Knowledge Society (AGESIC\footnote{https://www.gub.uy/agencia-gobierno-electronico-sociedad-informacionconocimiento/}) of Uruguay. 
However, these opportunities may be limited by data quality issues of this kind of scenarios \cite{Batini2016eGov}.}
\vero{In addition, such applications involve many services, processes and tasks, using varied data and metadata, for many purposes, and involving different users and roles. Thus, that DQ perceptions and expectations may be very different from a context to another.}

%\flavia{A case study developed for \vero{Uruguayan e-government agency,} AGESIC\footnote{\vero{AGESIC: e-Government Agency and Information and Knowledge Society, of Uruguay, https://www.gub.uy/agencia-gobierno-electronico-sociedad-informacionconocimiento/}}, presents a system called \textit{Citizen Claims System} inspired by a mobile application of the Municipality of the city of Montevideo in Uruguay. In this system, citizens submit claims, municipal officials attend the claims, and municipal managers make queries and reports on claims. In addition, some DQ problems arise from this interaction. For example, outdated citizen or claim information, duplicate claims, conflicting citizen data, etc. }
%
%\flavia{On the other hand, claims management involves several services, processes and data that are used for different purposes. In particular, the purpose for which data are used determines DQ. For instance, DQ expected by citizens is not the same that the expected by officials or managers. Since the application domain, the task at hand, the data used, and the user characteristics of the data determine the quality required for the data.}

%\cvero{To detail more, introducing an intuition of context and better illustrating the importance of context for DQM.}
Early approaches and surveys defined DQ as \textit{fitness for use} \cite{wang-strong} and showed the influence of %data 
context on DQ \cite{strong_data_1997}. %\flavia{However, models of DQ assessment have tended to ignore the impact of contextual quality on information use and decision outcomes \cite{shankaranarayanan2009}. In turn, the growing relevance of DQ has revealed the need for adequate measurement, using metrics for quantify DQ \cite{heinrich2007measure}.} 
\vero{For example,}
\flavia{%In \cite{pipino2002} is suggested that, 
for DQ assessment, a single set of DQ metrics is not} 
\vero{enough, and task-dependent DQ metrics, developed for specific application contexts, should be included \cite{pipino2002}.} 
%a solution, and it presents an approach that combines subjective and objective DQ metrics. The subjective DQ metrics are developed for specific application contexts. 
%They are called by the author task-dependent metrics. 
\vero{Enriching} %For instance, in 
DQ metrics definition \vero{with} context management would provide with: (i) flexibility, since they could adapt to context variations, (ii) generality, since they could include many particular context-dependent cases, and (iii) richness, leading to include more aspects to the metric. 

\vero{According to the literature, there is no single definition of context, rather it seems that context can be made up of several elements.
For instance, authors of \cite{pipino2002}}
\flavia{%Besides, they 
include the organization’s business rules, company and government regulations, and constraints provided by the database administrator, %Therefore, this implies that DQ metrics could depend not only on the task at hand, but also on other elements that make up the context of data. This confirms the contextual character of DQ measurement task.}
%
%\flavia{According to \cite{BolchiniCQST07}, context is something perceived by the user, 
while others \vero{include the} %assume that 
users %can be included in the context 
\cite{Dey01}, \vero{or the} %In the case of \cite{shankaranarayanan2009} the context is defined %, but in this case by
characteristics of the decision task and the decision-maker \cite{shankaranarayanan2009}. 
}
%Context is embedded in DQ management activities, for example when it is considered in the definition of DQ metrics, but it is generally not explicitly identified. 
%If context was defined and formalized, it could be used in DQ activities as an independent asset that impacts them.

%\flavia{In order to focus on a contextual DQ assessment and considering the aforementioned needs, it is necessary to define what is the context and formalize it, to later use it in DQ tasks.} 
%For instance, in DQ metrics definition, context management would provide with: (i) flexibility, since they could adapt to context variations, (ii) generality, since they could include many particular context-dependent cases, and (iii) richness, leading to include more aspects to the metric. 

Despite its recognized importance for DQ management \flavia{(DQM)}, context is typically taken as an abstract concept, which is not clearly defined nor formalized. In order to evidence this lack, we analyzed the state of the art on context definition for DQ management, with special attention to modeling and formalization aspects.

%\cvero{The motivation can be larger, including examples in industry, in e-government, in social initiatives... You can reuse motivation about Digital Government for this purpose.} \\
%\cpatrick{Indeed, having here a first idea of the running example would be nice}

\paragraph{Contributions} This paper presents a Systematic Literature Review (SLR) for investigating how\flavia{, when and where} context is taken into account in recent proposals for DQM. A SLR is a methodology for identifying, evaluating and interpreting all available research literature that is relevant to a particular research question or topic area. This methodology is not a conventional literature review, because while the latter just provides a high level summary of the literature in connected fields, the SLR provides a more complete, rigorous and reproducible framework for the literature study \cite{Kitchenham04}. 

\flavia{Specifically, we identify and analyze works dealing with context for DQM, and we classify them %identified research works based on a set of 
\vero{according to several} analysis criteria. %Examples of these analysis criteria are research domain and type of work. 
Besides, we identify contextual aspects of DQ tasks included at the stages of a DQM process. \vero{In addition, we describe} %In turn, a review of 
our findings about %the 
context definition %s is presented. In particular, we focus on the 
\vero{and} formalization \vero{as well as the} %of the context, 
identification of %the 
context components and their possible representations.}

\paragraph{Outline} The paper is organized as follows: %\cflavia{CHECK at the end} 
Section \ref{sec:background} provides background on the important notions used throughout this paper. 
Section \ref{sec:slr} presents the planning and execution of the SLR and \vero{describes its results.} %by classifying the selected papers according to several criteria.} 
Section \ref{sec:results} \vero{analyzes context proposals in terms of level of formalization, components and representation.} %we present an analysis of the proposed contexts. 
%In 
Section \ref{sec:examples}, \vero{details} %we describe 
the proposals for the formalization of context, %found with the SLR, through an example. 
\vero{and} %In 
Section \ref{sec:answers} %we 
answers the research questions \vero{of} %, proposed for 
the SLR. 
Finally, Section \ref{sec:conclusions} concludes and presents open research questions. 

The bibliography referenced in this work is classified into 2 groups: 
\vero{(i) the general bibliography, and the works selected by the SRL.}
%(in black) %(in blue)
 
%In black, the bibliography consulted taking into account the interest areas and, in blue, the bibliography found using the SLR.

%% file: _background.tex
\section{Background}
\label{sec:background}
This section presents background concepts about Data Quality% (DQ)
, Context and Systematic Literature Review. % (SLR).

\subsection{Data Quality}
DQ is defined as \textit{fitness for use} and is widely recognized to be multidimensional \cite{wang-strong}. \textit{DQ dimensions} express the characteristics that data must have, such as its correctness, completeness and consistency. In the literature, there is no agreement on the set of the dimensions characterizing DQ nor on their meanings \cite{scannapieco2002data}. This subsection lists the most representative ones, to be included in our review.

In a large industrial survey, Wang and Strong inventoried 175 quality attributes used by data consumers \cite{wang-strong} and %they 
organized them in 15 quality dimensions \cite{strong_data_1997}, namely, \textit{accuracy, objectivity, believability, reputation, accessibility, access security, relevancy, value-added, timeliness, completeness, amount of data, interpretability, ease of understanding, concise representation, consistent representation}. Such dimensions were also organized in 4 DQ categories: Intrinsic DQ, Accessibility DQ, Contextual DQ and Representational DQ. The ISO/IEC 25012 Standard \cite{25012-Stand} also proposes 15 quality dimensions (called characteristics), namely, \textit{accuracy, completeness, consistency, credibility, currentness, accessibility, compliance, confidentiality, efficiency, precision, traceability, understandability, availability, portability, recoverability}. These dimensions are reused by Merino et al. in the context of big data \cite{PS3}. In addition, the term \textit{data veracity} is widely used in Big Data applications, ambiguously relating to several DQ dimensions.

\vero{While DQ dimensions are used to express the characteristics that data must satisfy, \textit{DQ metrics} provide quantitative means for assessing to what extent these characteristics are satisfied.}
Pipino et al. in \cite{pipino2002} \vero{argue} %mention 
that it is difficult to define which aspects of a DQ dimension are related to a specific application in a company. However, once this task is done, proposing DQ metrics associated with these dimensions is easier. %In turn, usable DQ metrics are required to know how good is the quality of the data in the company. Taking this into account, the 
\vero{Their} research addresses DQ assessment based on \vero{task-dependent and -independent} %subjective and objective
DQ metrics. According to the authors, companies must consider the subjective perceptions of the %data 
users, and the objective measurements based on the used data. 

%Thus, while DQ dimensions are used to express the characteristics that data must satisfy, DQ metrics provide quantitative means for assessing to what extent these characteristics are satisfied. 
\vero{Also,} %At the same time, 
the term %``
\textit{DQ attribute} %" 
is often used in DQ literature, but ambiguously, as it may refer to DQ dimensions, to specific aspects of DQ dimensions, or even to DQ metrics. 

\vero{Concerning} %In relation to 
DQ tasks, in particular measuring, evaluating and improving tasks, it is not trivial to carry them out in a company\vero{, since} %. Since 
each of these tasks %involves a set of stages that 
generally depends on the complexity of the application domain. 
\vero{Typically, these tasks are organized and} %These stages are
carried out following a process, \vero{that is part of a} %also called 
methodology. %\citePS{PS16} defines 
A DQ methodology \vero{is defined} as a set of guidelines and techniques whose starting point is information related to a certain \vero{area} %reality 
of interest \vero{\citePS{PS16}}. 
%Bibliography suggests 
Several DQ methodologies \vero{have been proposed in the literature}, in particular, \citePS{PS16} compares 13 of them, \vero{tailored for DQ} %for information quality 
assessment and improvement. 

\vero{Finally,} %As we can see, 
in the literature we find the terms \textit{Data Quality} and \textit{Information Quality} interchangeably, as for example in \cite{pipino2002}. However, Batini and Scannapieco \cite{batini2016Book} use \vero{DQ} %the term \textit{data quality} 
when referring to dimensions, models, and techniques strictly related to structured data, while, in all other cases, they use %the term \textit
{information quality}. 
In DQ literature, there is no consensus on whether there is or not a difference between \vero{DQ} %data quality 
and information quality. Since, the difference is established or not, by the researchers of each proposal.

\subsection{Context}

In a tentative to understand context, Bazire and Brézillon \cite{DBLP:conf/context/BazireB05} analyzed a corpus of 150 definitions in different domains of cognitive sciences and close disciplines, and  build a model of context representing the components of a situation and the different relations between those components. As pointed out by Dey \cite{Dey01}, context is a poorly used source of information in computing environments, resulting in an impoverished understanding of what context is and how it can be used. Dey provided operational definitions of context and context-aware computing: context is a general term used to capture any information that can be used to characterize the situations of an entity, a system being context-aware if it uses context to provide relevant information and/or services to the user, where relevancy depends on the user’s task. 

Other concepts are used as synonymous of context. For instance, \textit{preferences} and \textit{data tailoring} appear as related concepts (see \cite{BolchiniCQST07, SerraM17} for a literature review). However, in some cases context-dependent preferences are also proposed \cite{AbbarMokrane,KostasPitoura,ciaccia2011}. Regardless of whether they represent the same thing, they are concepts that are strongly related. Profile is another concept that slightly differ from context. Nevertheless, when they are used jointly, the relationship between them remains often unclear \cite{AbbarMokrane}.

In a research on ubiquitous computing \cite{Abowd2000}, the authors explore context-aware computing. They point out that context is more than position and identity, and consider that although a complete definition of context is illusive, it is necessary to define a minimum set of context components\vero{, answering to} %: 
who, what, where, when and why \vero{questions}. But this is not the only important thing, since related to the context definition is the question of how to represent it.

\subsection{\vero{Context for Data Quality}}
%context in DQ
DQ assessment is a challenging task that requires data context and demands a great domain knowledge \cite{mylavarapu2020}. The importance and influence of data context in DQ has been stated many years ago \cite{wang-strong}, and is widely accepted. Early works only state whether a DQ dimension depends on context or not \cite{strong_data_1997}. However, \vero{since} %as early as 
2006, \vero{shankaranarayanan and Cai} %in 
\cite{shankaranarayanan2006} %it is 
pointed out that contextual factors have not been explicitly examined in data quality literature, and %the authors 
recognized the contextual or subjective nature of data quality evaluation. \vero{In addition,} %As well as, 
in 2009\vero{, Watts et al.} %the authors of
\cite{shankaranarayanan2009} \vero{argue} %mention 
that contextual assessments can be as important as objective quality indicators, because they can affect the information used for decision making tasks. Most recently, in 2019, \vero{Catania et al.} %\citePS{PS39} 
suggest that context information can help in interpreting users' needs in Linked Data. While in 2020, \vero{Davoudian and Liu} %the authors of 
\citePS{PS45} claim that traditional DQ tasks do not take into account Big Data characteristics, since \vero{the context of use is appropriate} for Big Data quality analysis. % must be appropriate for the context of use.

The need to consider the context for DQ assessment persists over time and for different application domains. Nevertheless, even today the literature lacks of a concise and globally accepted definition for context in DQ \citePS{PS25}. Therefore, this present work aims at investigating whether context is used and how \vero{it} is used in recent proposals for DQ management.

\subsection{Systematic Literature Review}

SLR \cite{Kitchenham04} is a methodology aiming at identifying, evaluating and interpreting all available research relevant to a particular research question or topic area. It starts by defining a review protocol that specifies the research questions being addressed and the methods that will be used to perform the review. These research questions determine the criteria to assess potential relevant data to answer the questions. Relevant data is extracted from scientific works, called \textit{primary studies} (\textit{PS} for short), that were found with the SLR. According to \vero{Kitchenham} \cite{Kitchenham04}, SRL methodology and produced documents (protocol, questions, etc.) ensure the completeness, rigor and reproducibility of the review.

As laid down in \cite{Kitchenham04}, a SLR is performed in three stages: (i) planning, (ii) conducting and (iii) reporting. At the first stage, the review objectives and the research questions are defined. Based on the research questions, the next stage consists in deriving search strings, selecting digital libraries and defining inclusion and exclusion criteria. The execution of search strings in search engines of digital libraries results in a set of PS. The last stage reports SLR results.

%% file: _slr.tex
\section{SLR Application}
\label{sec:slr}

A SLR is executed to obtain an overview of existing research on the use of Contexts in Data Quality. 
In this section, we describe the planning and conducting of the SLR methodology applied to our research problem, and we present quantitative and qualitative results. More detailed findings, including the description of some proposals, are presented in Section \ref{sec:results}. 

\subsection{Planning the review}

At this stage, the objectives of the investigation, and the research questions that arise from these objectives, are defined.

\paragraph{Review Objectives:}
We apply a SLR to know current evidence in the relationship between \textit{Context} and \textit{Data Quality} areas. In short, we want to identify any gap at the intersection of these areas in order to detect new research lines.

\paragraph{Research Questions:}
To define the research questions, the most important DQ concepts (DQ model, DQ dimension and DQ metric) are taken into account. In particular, it is important to know if DQ models have been defined considering context and how context is represented in such models. 
In turn, we search whether context participates in the definition of DQ dimensions and their DQ metrics, and more generally, how context relates to DQ concepts at different stages of DQ management. For all this, the following research questions are raised:
\begin{itemize}
    \item \textbf{RQ1}: How is context used in data quality models?
    \item \textbf{RQ2}: How is context used within quality metrics for the main data quality dimensions?
    \item \textbf{RQ3}: How is context used within data quality concepts?
\end{itemize}
Note that RQ3 is intentionally more inclusive, in order to get works dealing with DQ and context but not mentionning specific DQ models, DQ dimensions or DQ metrics.

\subsection{Conducting the review}
The SLR execution is carried out following a series steps: search strings definition, selection of the digital libraries, definition of the inclusion and exclusion criteria, and selection of the primary studies. All of them are described below.

\paragraph{Search Strings:}
To create the search strings we firstly extract the keywords of the research questions, namely: \textit{context}, \textit{data quality model}, \textit{quality metric}, \textit{data quality dimension}, \textit{quality concept} and \textit{data quality}. The latter two are the decomposition of a more complex keyword, \textit{data quality concept}, appearing in RQ3. Then, to perform a thorough search, some keywords are complemented or detailed with alternative terms as follows (each of them is discussed and argued in Section  \ref{sec:background}): 

\begin{itemize}
    \item \textit{context} is complemented with alternative terms \textit{data tailoring} and \textit{preference}.%, as discussed in Section \ref{sec:background}. 
    
    \item \textit{data quality dimension is refined by} the DQ dimensions. %listed in Section \ref{sec:background}. 
    Note that, contrarily to dimensions, metrics are not refined. Indeed, unlike dimensions, metrics are usually not managed with names in DQ literature, since they are defined specifically for each measurement process. 
    
    \item \textit{quality concept} is refined by the most important DQ 
    concepts: \textit{quality dimension, quality metric} and \textit{quality attribute}.%, also described in Section \ref{sec:background}.

    \item \textit{data quality} is complemented with alternative term \textit{information quality}.%, as argued in Section \ref{sec:background}.
\end{itemize}

\begin{table}[t] 
 \begin{center}
 \begin{tabularx}{\linewidth}{|>{\raggedright\setlength\hsize{0.19\hsize}}X| >{\setlength\hsize{0.81\hsize}}X|}
 \hline
 \makebox[2cm][c]{\textbf{Keywords}} & \makebox[10cm][c]{\textbf{Alternative terms}} \\
 \hline \hline
 context 
     & preference, ``data tailoring'' 
 \\ \hline

 data quality model 
     &  \\ \hline

 quality metric 
     &  \\ \hline

 data quality  dimensions
     & ``data believability", ``data accuracy", ``data objectivity", ``data reputation", ``data value-added'', ``data relevancy'', ``data timeliness'', ``data completeness'', ``appropriate amount of data'', ``data interpretability'', ``data ease'', ``data understanding'', ``data representational consistency'', ``data concise representation'', ``data credibility'', ``data currentness'', ``data veracity'', ``data accessibility'', ``data compliance'', ``data confidentiality'', ``data efficiency'', ``data precision'', ``data traceability'', ``data understandability'', ``data availability'', ``data portability'', ``data recoverability'' 
     \\ \hline

 quality concept 
     & ``quality dimension'', ``quality metric'', ``quality attribute'' \\ \hline

 data quality  
     & ``information quality'' \\ \hline

 \end{tabularx}
 \caption{Alternative terms for each keywords.}
 \label{tab:keywords}
 \end{center}
 \end{table}

Then, for each research question, a search string is built as a conjunction of keywords (AND connector). The alternative terms are joined with the keywords by means of disjunctions (OR connectors). We can see the alternative terms of each keywords in Table \ref{tab:keywords}.
Since the RQ2 is very comprehensive the associated search string is too long to be supported by search engines of some digital libraries, as it should list 27 DQ dimensions. To solve this problem, we cut the  search string in 7 smaller ones, each containing a subset of the DQ dimensions. The resulting search strings are:

\begin{itemize}
    \item \textbf{SS1}: \textbf{(}context OR preference OR ``data tailoring''\textbf{)} AND \textbf{(}``data quality model''\textbf{)}
    
    \item \textbf{SS2}: \textbf{(}context OR preference OR ``data tailoring''\textbf{)} AND \textbf{(}``quality metric''\textbf{)} AND \textbf{(}``data believability'' OR ``data accuracy'' OR ``data objectivity'' OR ``data reputation''\textbf{)}
    
    \item \textbf{SS3}: \textbf{(}Context OR preference OR ``data tailoring''\textbf{)} AND \textbf{(}``quality metric''\textbf{)} AND \textbf{(}``data value-added'' OR ``data relevancy'' OR ``data timeliness'' OR ``data completeness'' OR ``appropriate amount of data''\textbf{)}
    
    \item \textbf{SS4}: \textbf{(}Context OR preference OR ``data tailoring''\textbf{)} AND \textbf{(}``quality metric''\textbf{)} AND \textbf{(}``data interpretability'' OR ``data ease'' OR ``data understanding'' OR ``data representational consistency'' OR ``data concise representation''\textbf{)}
    
    \item \textbf{SS5}: \textbf{(}Context OR preference OR ``data tailoring''\textbf{)} AND \textbf{(}``quality metric''\textbf{)} AND \textbf{(}``data credibility'' OR ``data currentness'' OR ``data veracity''\textbf{)}

    \item \textbf{SS6}: \textbf{(}Context OR preference OR ``data tailoring''\textbf{)} AND \textbf{(}``quality metric''\textbf{)} AND \textbf{(}``data accessibility'' OR ``data compliance'' OR ``data confidentiality'' OR ``data efficiency''\textbf{)}
    
    \item \textbf{SS7}: \textbf{(}Context OR preference OR ``data tailoring''\textbf{)} AND \textbf{(}``quality metric''\textbf{)} AND \textbf{(}``data precision'' OR ``data traceability'' OR ``data understandability''\textbf{)}
    
    \item \textbf{SS8}: \textbf{(}Context OR preference OR ``data tailoring''\textbf{)} AND \textbf{(}``quality metric''\textbf{)} AND \textbf{(}``data availability'' OR ``data portability'' OR ``data recoverability''\textbf{)}
    
    \item \textbf{SS9}: \textbf{(}Context OR preference OR ``data tailoring''\textbf{)} AND \textbf{(}``data quality'' OR ``information quality'' \textbf{)} AND \textbf{(}``quality metric'' OR ``quality dimension'' OR ``quality attributes''\textbf{)}
\end{itemize}

%\begin{figure}[t]
%\centering
%\captionsetup{justification=centering}
%\includegraphics[scale=0.7]{Figures/fig-search-strings.png}
%\caption{Search strings} 
%\label{fig:searchStrings}
%\end{figure}

%%%%%%%%%%%%%%%%%%%%%%%%%%%%%%%%%%%%%%%%%

\paragraph{Digital Libraries:}
We select digital libraries trying to cover the most important venues in the database domain, but having the least possible overlap among them, thus reducing the number of duplicated PS. In addition, the selected digital libraries should allow for search in both titles and abstracts. We initially avoided to use the search engines of DBLP and Google Scholar because most of the returned PS are already returned by the other digital libraries. As well, Google Scholar indexes many other types of documents (e.g. technical reports or slides) producing noise within search results and DBLP does not support search inside abstracts, loosing many relevant PS. However, we are aware that some important venues have online proceedings, in particular VLDB, EDBT and ICDT. So, we complete the SLR with a specific search in Google Scholar, but restricted to VLDB, EDBT and ICDT.

The selected digital libraries 
are the following:
\begin{itemize}
    \item ACM digital library (http://dl.acm.org)
    \item IEEE Xplore (http://ieeexplore.org)
    \item Science Direct (http://www.sciencedirect.com)
    \item Springer LNCS (http://www.springer.com)
    \item Google Scholar (https://scholar.google.com) restricted to 3 
    venues.
\end{itemize}

\paragraph{Inclusion and Exclusion Criteria:}
To restrict the PS returned by each digital library, we apply a set of inclusion and exclusion criteria. The former are automatically applied by digital libraries, while the latter are considered later, while reading abstracts or full texts.

The considered inclusion criteria are: (i) PS are published \flavia{in the period 2010-2020} inclusive. This time interval is large enough to ensure getting recent PS. (ii) PS are written in English and (iii) are published in PDF format. (iv) The set of PS only includes articles and book chapters.  

\flavia{Using exclusion criteria, we selected PS} if: (i) full text is written in another language (even if the abstract is in English), (ii) they are published in non pair-reviewed venues, (iii) data quality is not addressed or it is addressed superficially, (iv) context is not addressed. These criteria allow to discard PS addressing other subjects even if containing the good keywords.

% \begin{table}[H]
% \begin{center}
% \begin{tabularx}{\linewidth}{|l|X|}
% \hline
% \makebox[5cm][c]{\textbf{Inclusion criteria}} & \makebox[7cm][c]{\textbf{Exclusion criteria}} \\
% \hline \hline
% published since 2010 inclusive 
%     & abstract in English but full text in another language \\ \hline
% in English language            
%     & quality is addressed but not data quality (e.g. context quality)  \\ \hline
% must be an article or book chapter 
%     & data quality is addressed but superficially  \\ \hline
% in PDF format 
%     & context is not addressed \\ \hline
% \end{tabularx}
% \caption{Inclusion and exclusion criteria.}
% \label{tab:incExc-crit}
% \end{center}
% \end{table}

% \begin{figure}[t]
% \centering
% \captionsetup{justification=centering}
% \includegraphics[scale=0.45]{Figures/fig-selection-ps.png}
% \caption{Outline of the selection process and quantity of PS at each step.} 
% \label{fig:selectionProcess}
% \end{figure}

\paragraph{Selection of Primary Studies:}
The selection process has 5 steps. They are listed below, highlighting the number of PS output by each step, which are summarized in Table \ref{tab:resPerDL}.
\begin{itemize}
    \item \textbf{Step 1: Execution of search strings.} 2898 PS were returned by the 5 digital libraries in response to search strings and inclusion criteria. 
    \item \textbf{Step 2: Duplicate elimination.} Some PS were returned by several search strings within the same digital library, and many PS from Google Scholar were also returned by other libraries. Duplicate elimination resulted in 1955 PS. 
    \item \textbf{Step 3: Selection by relevance.} 279 PS were selected by relevance, reading the title and abstract of each PS.
    \item \textbf{Step 4: Selection by full text.} A final selection, by a careful reading of full text and application of exclusion criteria, resulted in 54 PS.
    \item \textbf{Step 5: Selection by references.} Finally, the references of the 54 selected articles are reviewed. Respecting the inclusion and exclusion criteria, 4 additional articles were selected, obtaining a total of 58 PS.
\end{itemize}

\begin{table}[t]
\begin{center}
\begin{tabular}{|l|c|c|c|c|c| }
\hline
\textbf{Digital Library} & \textbf{Step 1} & \textbf{Step 2} & \textbf{Step 3} & \textbf{Step 4} & \textbf{Step 5} \\ 
\hline
ACM             &  506 &      & 120 & 15 & \\
IEEE            &   11 &      &   3 &  0 & \\
ScienceDirect   & 1027 &      &  48 &  9 & \\
Springer        & 1287 &      & 107 & 29 & \\
Google Scholar (VLDB, EDBT, ICDT)    
                &   67 &      &   1 &  1 & \\\hline
From references & & & & & 4 \\
\hline
{\textbf{Total}} & 2898 & 1955 & 279 & \textbf{54} & \textbf{58} \\ 
\hline   

\end{tabular}
\caption{Amount of selected PS by step and Digital Library}
\label{tab:resPerDL}
\end{center}
\end{table}

\begin{figure}[H]
\centering
\captionsetup{justification=centering}
\includegraphics[scale=0.55]{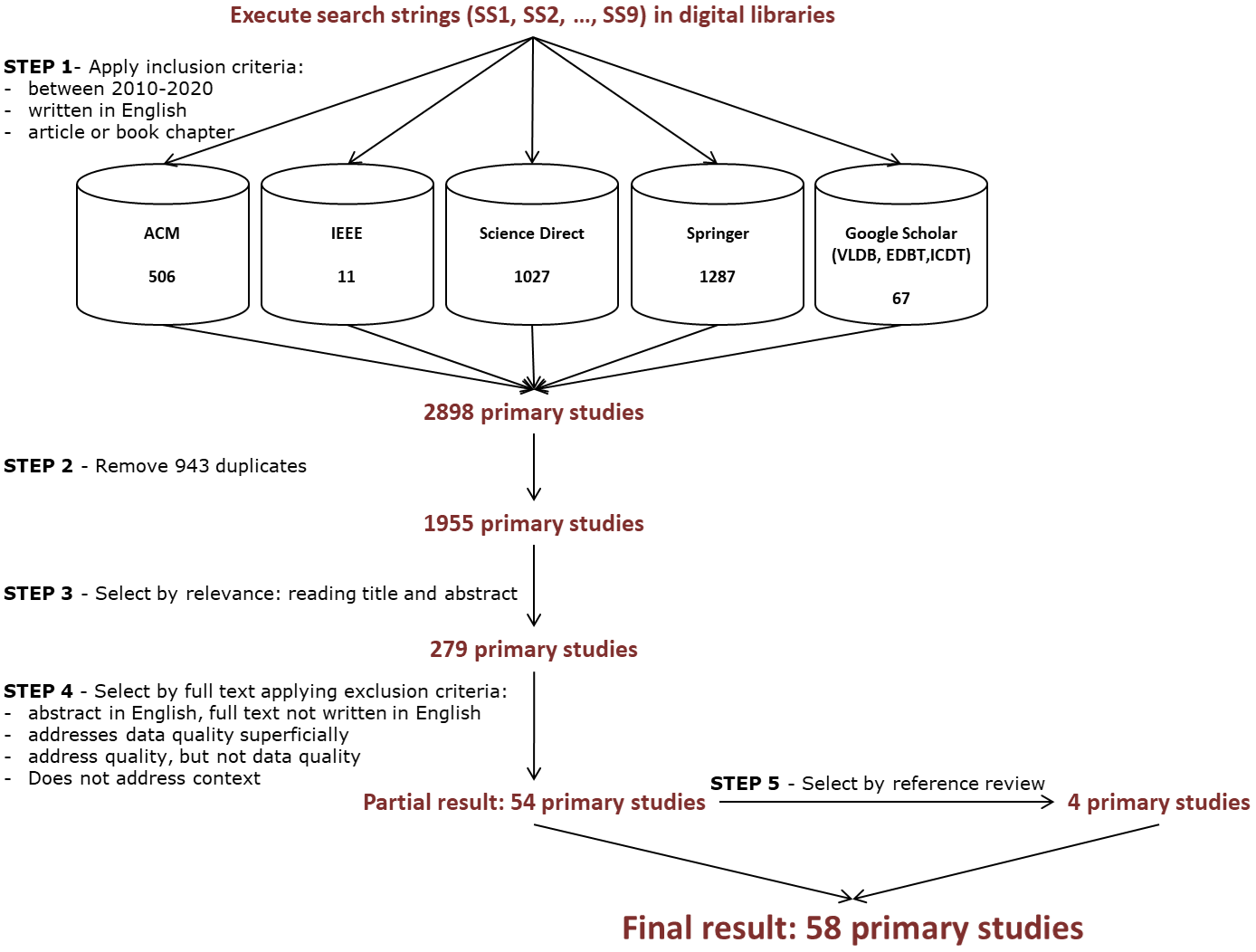}
\caption{Primary studies selection process.}
\label{fig:SLRprocess}
\end{figure}

According to \cite{Kitchenham04} this methodology is reproducible. However, there are some exceptions as the methodology depends on external factors. In particular, we experienced some problems with the ACM library, which changed its search engine in the middle of the SLR process. Indeed, our first search returned very few PS, while a second search some month later, returned hundreds of PS. This evidences, that the reproducibility of the methodology, as described in \cite{Kitchenham04}, is impacted by the changes in digital libraries. The complete selection process is shown in Figure \ref{fig:SLRprocess}.

\subsection{Review results}
\label{sec:slr:results}

In this section we present and quantify the main results of the review by analyzing the selected PS according to different criteria. From now on, we term PS is used for referring to selected PS. 

\begin{figure}[hbt]
\centering
\captionsetup{justification=centering}
\includegraphics[scale=0.55]{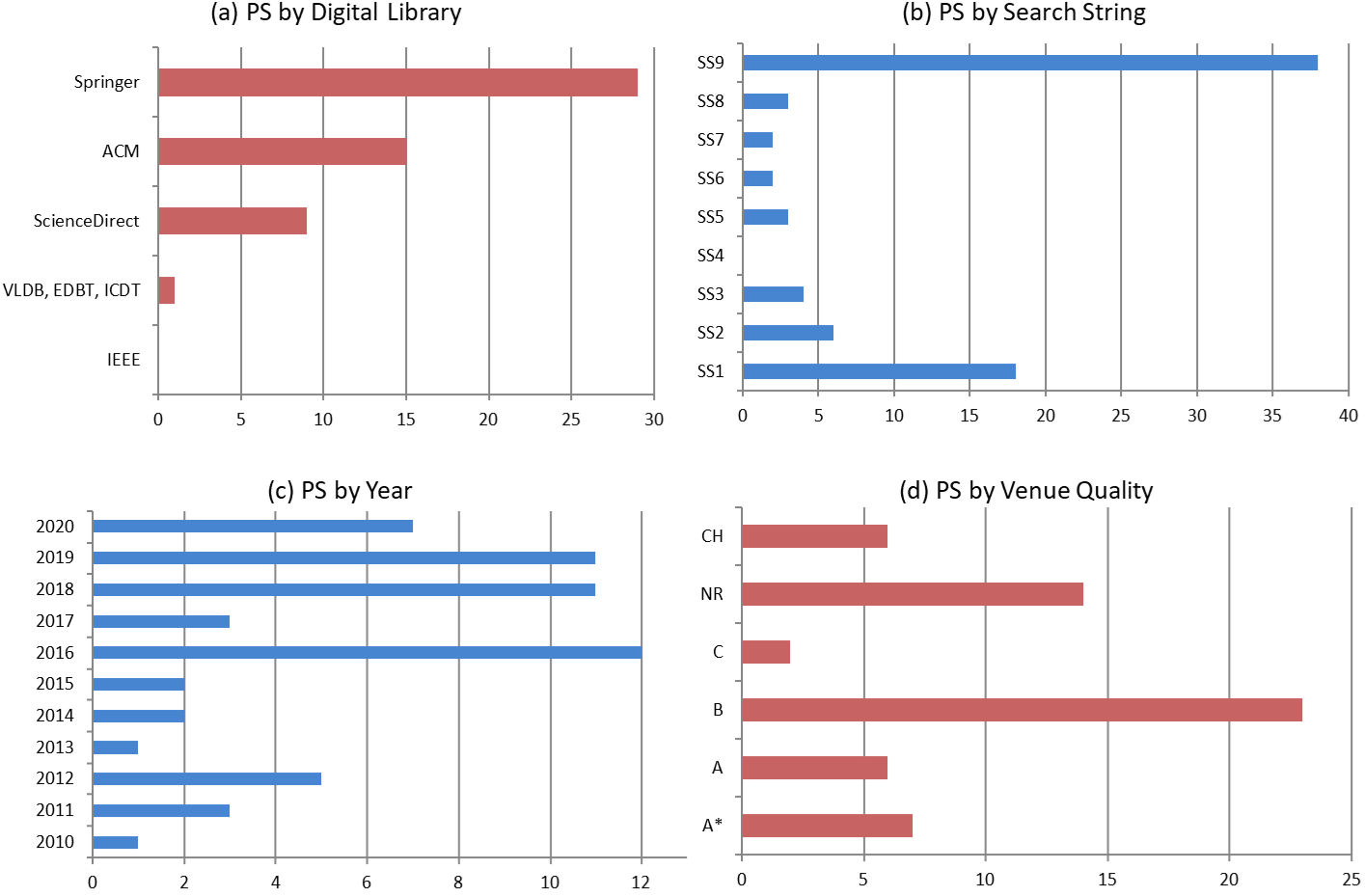}
\caption{Total of PS by digital library (a), search string (b), year of publication (c) and venue quality (d).
} 
\label{fig:SLRresults}
\end{figure}

Firstly, for complementing Table \ref{tab:resPerDL}, Fig. \ref{fig:SLRresults} shows the distribution of selected PS by (a) digital library, (b) search string, (c) year of publication and (d) venue quality. We remark that more than a half of the selected PS were retrieved with SS9, while none with SS4, and most of selected PS come from Springer and ACM, while none from IEEE. Interestingly, the number of published papers dealing with the use of context for DQ increased from 2016, except for 2017. Selected PS are classified according to the quality of venues, in accordance with rankings and metrics of Scopus\footnote{SCOPUS portal. https://www.scopus.com/sources (accessed December 2020)} for journals  and Core\footnote{CORE portal. http://portal.core.edu.au/conf-ranks/ (accessed December 2020)} for conferences.
We use the quality labels (A*, A, B, C) proposed by Core, which correspond to Scopus relative rankings of [94\%-100\%], [84\%-94\%), [54\%-84\%) and [13\%-43\%), respectively. NR is used for referring to not ranked PS and CH for referring to book chapters. We remark that only 13 PS (22\%) are ranked A* or A, and interestingly, 11 of them were published in or after 2016. In addition, 23 PS are ranked B (including 11 from the Int. J. on Data and Information Quality) and 14 PS are not ranked. Concerning venue type, 31 PS come from conference proceedings, 21 from journals and 6 from book chapters.

The selected PS are classified according to the following analysis axes: 
\begin{itemize}
    \item Type of work. We classify the PS according to the type of proposal. We find several approaches: framework, model, methodology, analysis, case study, metrics and architecture. A lot of PS address more than one approach. In particular, this is the case of PS that propose a framework. These generally include a model and/or a methodology, among others. \flavia{We focus on the proposals that contribute the most to our research}.
    
    \item Research domain. We consider the research domain addressed in the PS. Firstly, we highlight PS that only address data or information quality \flavia{and we call it \textit{only DQ}}. These PS address, implicitly or explicitly, contextual aspects of DQ. On the other hand, we have other PS that include contextual DQ in their researches. However, in these cases the main research domain is one of the following: Big Data, Decision Making, Linked Data, Internet of Thing, Data Mining, e-Government, Open Data, Machine Learning or Data Integration. \flavia{These domains use DQ or are exploited to improve DQ.}
    
    \item DQ process stages. We analyze which are the stages, of the DQ process, addressed in the PS. As well, we identify the most important DQ process stages, according to the bibliography.
    
    \item DQ dimensions. Some authors affirm that there are contextual DQ dimensions. We classify the PS according to this proposal. That is, we are interested in identifying how many PS address the contextual DQ dimensions approach.
    
    \item DQ metrics. In this case, the approach is similar to the previous classification. Then, we identify the PS that define, use or analyze contextual DQ metrics.

    \item Data model. We also have a classification according to the data model used in the selected PS. In this way, we find PS with structured, semi-structured, unstructured and mixed \vero{data models}. In some PS data are not specified or PS are general, i.e., not restricted to a data type or model. In both cases, PS are classified as N/A.
    
    \item Case study: The PS are also classified by the type of data used in the case study presented to evaluate the proposal. The case studies can use real or artificial data. In some PS a case study is not specified (N/A).
    
\end{itemize}

\begin{figure}[hbt]
\centering
\captionsetup{justification=centering}
\includegraphics[scale=0.55]{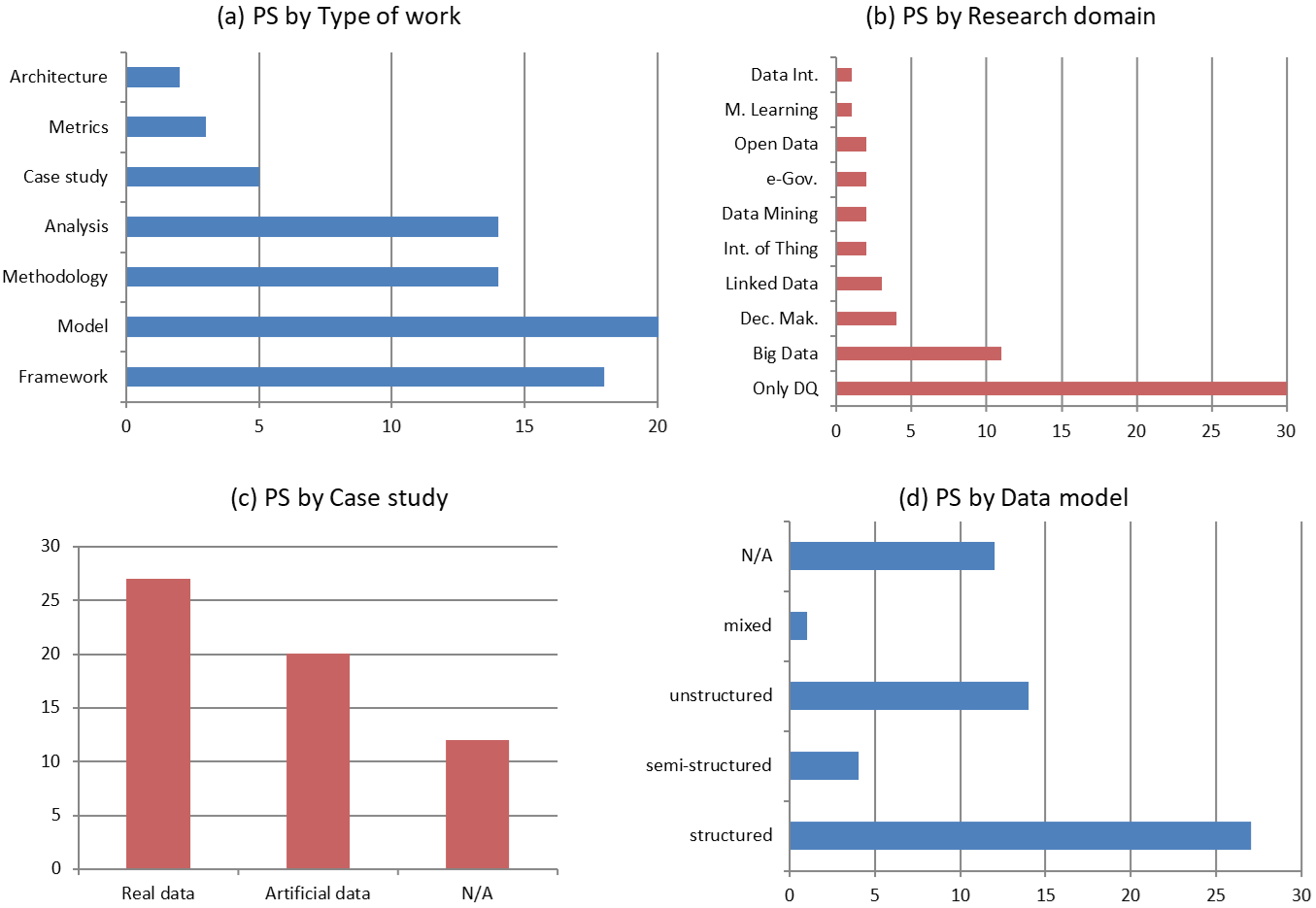}
\caption{Total of PS by type of work (a), research domain (b), case study (c), and data model (d).}
\label{fig:axis}
\end{figure}

Next paragraphs present the classification results for each analysis axis. Firstly, each paper is introduced according to its type of work. Then, the salient facts are commented for the other analysis axes.

\paragraph{Type of work} In this classification we present the different types of works found in the selected PS. When PS have several contributions, we focus on the main proposals in order to indicate the leading types of work.

The most frequent type of work, as evidenced in Fig. \ref{fig:axis}(a), is the \textit{modeling}, since several PS coincide in proposing a DQ model \citePS{PS2,PS3,PS13,PS18,PS20,PS27,PS28,PS33,PS34,PS59}. In particular, \citePS{PS59} presents domain-specific characteristics of DQ, while \citePS{PS33} extends this work by proposing DQ models. In addition, in \citePS{PS15}, the authors discuss models proposed for utility, discovering that DQ dimensions and DQ metrics are deeply influenced by utility. This proposal considers the usage of data and the relevance of processes that adopt it in the measurement. In turn, \citePS{PS32} presents a model to identify opportunities for increasing monetary and non-monetary benefits by improved DQ. In other matters, we identify other kind of models, all for DQ assessment and motivated by the idea that DQ is context-dependent. For instance, the authors of \citePS{PS52} present a decision model to facilitate the description of business rules, and in \citePS{PS11,PS41,PS57} a model of context is developed. As well, there are PS that address the impact of poor DQ and propose improvement models, specifically \citePS{PS37} presents a machine learning model and \citePS{PS46} a neural networks model.

\flavia{On the other hand, \citePS{PS7} presents a model of the asset management data infrastructure and \citePS{PS12} a Bayesian Network model that shows how DQ can be improved while satisfying budget constraints. In the case of \citePS{PS19}, it presents a formal multidimensional model on which is applied a rule-based approach to DQ assessment, and \citePS{PS24} provides a cost and value model for issues related to DQ. Finally, \citePS{PS39} models context for source selection in Linked Data, and \citePS{PS44} proposes a conceptual model where the relationships among quality characteristics of e-government websites and users’ perceptions are represented.} 

Another category concern \textit{frameworks} for assessing data or information quality \citePS{PS8,PS13,PS14,PS19,PS23,PS24,PS25,PS31,PS34,PS39,PS47,PS49,PS55,PS56,PS58}. 
For instance, in \citePS{PS39} the authors use a framework called \textit{Luzzu}\cite{Luzzu} for linked DQ assessment. Additionally, \citePS{PS19} provides a framework for context-aware Data Warehouse quality assessment. Other proposals that are task-specific, as \citePS{PS12} that presents a framework for discovering dependencies between DQ dimensions, or \citePS{PS54} that proposes a framework that encapsulates data cleaning procedures. Also in \citePS{PS55} we find an ontology-based quality framework for data integration. 
On the other hand, in some frameworks DQ is not the goal, but it is a relevant element of the proposal. For example, in \citePS{PS44} a framework of citizens’ adoption of e-government services is presented. Furthermore, an alternative angle is presented in \citePS{PS24}, since the authors develop a framework in the form of a guideline to manage DQ costs. They define DQ cost as financial impacts caused by DQ problems and resources required to achieve a specified level of DQ. Finally, the framework in \citePS{PS31} describes the implementation of a system that acquires tweets in real time, and applying DQ metrics to measure the quality of such tweets. These metrics are defined formally.

Among the works classified as \textit{analysis}, we find reviews and surveys. We consider them together because the goal of both is the same, to analyze the state of the art. For instance, \citePS{PS1} identifies the core topics and themes that define the identity of DQ research, whereas \citePS{PS26} analyzes the main concepts and activities of DQ. In the case of \citePS{PS6}, the authors investigate DQ in the context of Internet of Thing. In \citePS{PS9}, besides presenting a DQ methodology, the authors review DQ dimensions and other methodologies to assess DQ. In addition, some PS relate DQ with a particular domain, for example, Big Data \citePS{PS17,PS45}, Data Warehouses \citePS{PS25} and Open Data \citePS{PS30}. Moreover, \citePS{PS48} studies the mapping between metadata and DQ issues, in particular, the connection between metadata (such as count of rows, count of nulls, count of values and count of value pattern), and data errors. In other matters, \citePS{PS21,PS53} apply a systematic literature review, the former to find works that consider DQ requirements during the process of developing an information system, the latter presents our preliminary results of review on DQ and context. In addition, in \citePS{PS16} the authors compare methodologies that have been proposed in the literature for information quality assessment and improvement.

On the other hand, we find PS focused on \textit{methodologies}. The most of these work are focused on methodologies for data or information quality, specifically considering evaluation and improvement tasks \citePS{PS4,PS9,PS10,PS13,PS16,PS23,PS35,PS36,PS38,PS51}. 
As seen before, \citePS{PS13} describes a DQ framework, but it also includes a DQ methodology. \citePS{PS23} employs a particular methodology, called six-sigma \cite{linderman2003six}, to define critical factors to quality, measure current quality level, analyze deficiencies in information and identify the root causes of poor information. With the same purpose, \citePS{PS4} considers a methodology to build a DQ adapter module, which selects the best configuration for the DQ assessment. In turn, a data-focused methodology, based on DQ actions, is used in \citePS{PS36} to get smart data and to obtain more value for the datasets. Additionally, with another focus, a methodology for selecting sources in the Linked Data domain is developed in \citePS{PS50}. In this process context-dependent DQ is taken into account, according to different DQ dimensions. 

We also find works that present \textit{case studies} focusing on the analysis of a particular problem. Firstly, \citePS{PS7} evaluates how data governance supports data-driven decision-making in asset management organizations. Also, it investigates the impact of data governance on DQ. Otherwise, the goal of \citePS{PS29,PS30} is to examine the quality of Open Data in public sector, while \citePS{PS31} dives into the problem of re-defining traditional DQ dimensions and DQ metrics in the Big Data domain. 

In the \textit{metrics} category, \citePS{PS22} investigates how metrics for the DQ dimensions completeness, validity, and currency can be aggregated to derive an indicator for the accuracy dimension. In \citePS{PS42}, a set of requirements for DQ metrics is presented. By last, \citePS{PS43} proposes a visual analytic approach that enables data analysts to utilize and customize quality metrics, in order to facilitate DQ assessment of their specific datasets.

In relation to works that focus on \textit{architectures} we have two proposals. \citePS{PS5} presents a system architecture that includes mechanisms for DQ assessment and security, while \citePS{PS35} provides a DQ assessment architecture that manage streaming data in a context-aware manner at different levels of granularity.

%By way of summary, we can see that frameworks are the most used by the researchers. Whereas in second place we find the models, following by methodologies and reviews. In Fig. \ref{fig:axis} (a), we have the complete distribution of values for the types of works. 

\paragraph{Research domain} 
\vero{Fig. \ref{fig:axis}(b) shows the distribution of the main research domains of the selected PS. Most of them only concern}
%Since the research is focused on data quality and context, most of the selected PS only address 
data/information quality (only DQ) \vero{but their proposals are not domain-specific.}  
However, we can see that many PS combine DQ with other research domains. For example, the Big Data domain, where the management of large volumes of data presents a clear need to incorporate DQ tasks \citePS{PS1,PS17,PS35} %. In addition, several PS analyze the need to incorporate data
\vero{and study} context into those DQ tasks \citePS{PS3,PS4,PS18,PS28,PS31,PS45}.
%On the other hand, Linked Data (LD) \citePS{PS39} and Decision Making (DMak) \citePS{PS2} are research domains that highlight the need to relate data quality and data context. However, \citePS{PS39,PS2}, are the only most recent ones regarding these research domains. With similar results, we have the Data Mining (DM), e-Government (e-gov), Internet of Thing (IT) and Open Data (OD) domains. Because, as well in these cases, data context is little considered in DQ assessment. The same happens for the Data Integration (DI) and Machine Learning (ML) domains.

\begin{figure}[t]
\centering
\captionsetup{justification=centering}
\includegraphics[scale=0.55]{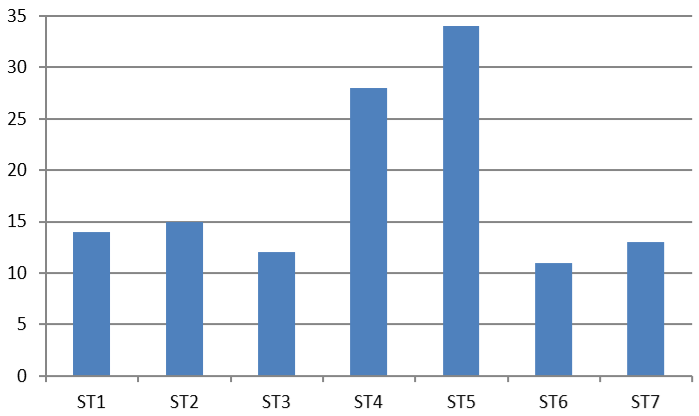}
\caption{Total of PS by DQ process stages.}  
\label{fig:DQproStages}
\end{figure}

\paragraph{DQ process stages}
We also analyze the PS according to the stages of the DQ process that are addressed in the proposal. As we mentioned in Section \ref{sec:background}, we consider a DQ process with 7 stages: ST1-Characterize Scenario, ST2-Analyze Objective Data, ST3-Define Strategy, ST4-Define Data Quality Model, ST5-Measure and Evaluate Data Quality, ST6-Determine Causes of Quality Problems, and finally, ST7-Define, Execute and Evaluate Action Plan. 
For this classification, we observe that some PS address more than one stage. In \vero{Fig. \ref{fig:DQproStages} we show these results.}
Measurement and evaluation of DQ, and the definition of a DQ model are the DQ process stages most addressed by the PS. The rest of the stages are considered in a similar way among the PS, in particular only 1 PS do not focus on any of these stages.

%los PS clasificados de acuerdo a cada etapa se pueden ver en la Tabla \ref{tab:ctxCompXstage} (PS by Context Components and DQ Process Stages)

\begin{figure}[t]
\centering
\captionsetup{justification=centering}
\includegraphics[scale=0.55]{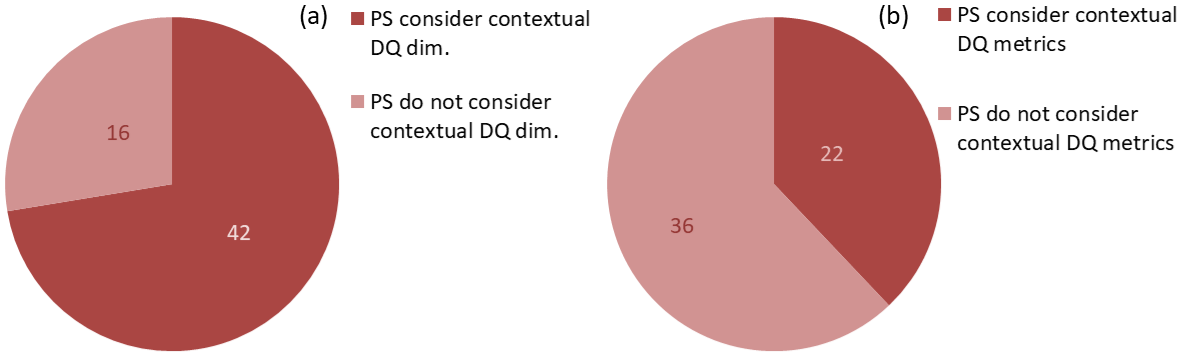}
\caption{Total of PS that consider contextual DQ dimensions (a) and contextual DQ metrics (b).}
\label{fig:ctxDim-Met}
\end{figure}

\begin{figure}[t]
\centering
\captionsetup{justification=centering}
\includegraphics[scale=0.70]{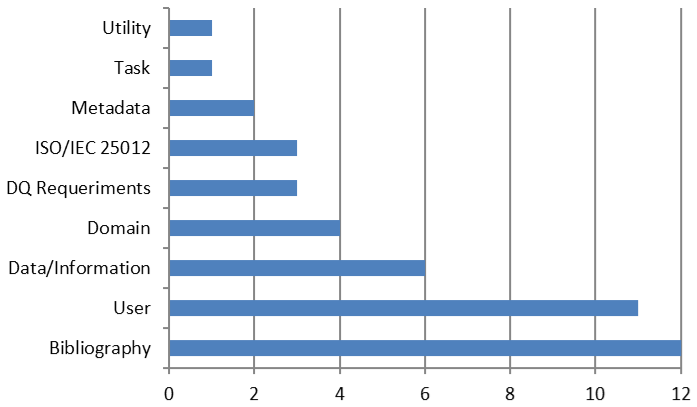}
\caption{Total of PS with contextual DQ dimensions according to different approaches.}  
\label{fig:DQdim}
\end{figure}

\paragraph{DQ dimensions} Several authors affirm that there is a set of DQ dimensions that are contextual, i.e., context dependent DQ dimensions. However, there is no agreement on what ensures the contextual characteristics of these DQ dimensions. Most authors rely on the bibliography to ensure that certain DQ dimensions are contextual \citePS{PS1,PS2,PS6,PS10,PS13,PS20,PS26,PS31,PS45,PS46,PS47,PS49}. For example, authors are based on reference bibliography of the DQ domain, such as \cite{wang-strong, Ge2007ARO}.
Furthermore, different aspects of the user (data profile, location, requirements, etc.) are also taken into account for contextualizing DQ dimensions \citePS{PS4,PS8,PS9,PS23,PS27,PS32,PS34,PS35,PS55,PS56,PS58}. In turn, data in use or information are also used to give context to DQ dimensions \citePS{PS12,PS14,PS15,PS19,PS38,PS39}, and according to some PS, DQ dimensions can be influenced by rules and constraints of the application domain \citePS{PS16,PS25,PS43,PS51}.
On the other hand, some authors affirm that DQ requirements condition DQ dimensiones \citePS{PS5,PS42,PS52}. But also, we can find PS where the contextual aspects of DQ dimensions are based on the ISO/IEC 25012 standard \footnote{https://iso25000.com/index.php/en/iso-25000-standards/iso-25012} \citePS{PS3,PS17,PS21}. To a lesser extent, authors claim that the contextual aspects of the DQ dimensions depend on metadata \citePS{PS22,PS48} and the task at hand \citePS{PS18}. As we saw before in \citePS{PS15}, from another perspective Batini and Scannapieco analyze models for utility-driven quality assessment. The authors underline that utility can help define superior measurements for DQ dimensions (e.g., for completeness and accuracy), that reflect DQ assessment in context. In Fig. \ref{fig:ctxDim-Met} we can see the total number of PS that consider contextual DQ dimensions. In turn, Fig. \ref{fig:DQdim} shows the quantity of PS for each of the approaches. Finally, it is not possible to find a single set of contextual DQ dimensions, since they vary among the different proposals. However, among the PS the DQ dimensions most commonly considered contextual are completeness, accuracy, consistency, relevance and timeliness. 

\paragraph{DQ metrics} In this classification, we consider PS that define, use or analyze contextual DQ metrics. Each contextual DQ metric takes into account different contextual aspects, since it strongly depends on the proposal. This means that DQ metrics are influenced by the type of data being measured and how they are measured, among others. According to Heinrich et al. \cite{Heinrich09}, DQ metrics need to be adaptable to the context of a particular application. In Fig. \ref{fig:ctxDim-Met}, we also show the quantity of PS that consider contextual DQ metrics.

\paragraph{Case study}
According to the Fig. \ref{fig:axis}(c), most of the PS use real data to develop their case studies. At the same time, we have some PS that do not apply (N/A) in this classification. The latter is so because these PS correspond especially to reviews and surveys. 

\paragraph{Data model}
Fig. \ref{fig:axis}(d) shows the distribution of the data models used in the selected PS. As we can see, most of the researches are based on structured data, especially using the relational schema. There are also a high number of PS that consider unstructured data, as CSV files or sensor data. On the other hand, there are an important number of PS that are general, i.e. they are not restricted to a data type or model. Thus, they are classified as N/A.
\\ \\
In the next section, we will present the proposed contexts. In particular, we will focus on presenting new classifications for the selected PS. This analysis will be based on the characteristics of the proposed contexts, as the level of formalization of context definition, the context components and the different context representations.

%% file: _results.tex
\section{Analysis of the proposed contexts}
\label{sec:results}

%In this section we analyze the definition and components of the context in the PS selected in the SLR. Specifically, we describe three main issues: the level of formalization of context definition, the various components composing the context, and the representation of those components.

%\cvero{For me, the following paragraphs are not necessary here. They can be mixed with the explanations of the corresponding subsections. The last sentence can remain.}

%In the case of context definition, we classify it in three levels of formalization: formal, no formal \cpatrick{I strongly suggest using informal rather than no formal} or implicit. 
%\cpatrick{clarify implicit: no definition (neither formal nor informal) is given}
%In particular, in Section \ref{sec:examples}, the formal context definitions are explained through an example. This example allows us to represent each of the elements of the formalization.

Many works affirm that DQ depends on the context, but what is the context? There are many proposals. For example, in \citePS{PS35}\citePS{PS23} the authors claim that DQ assessment depends on the user, i.e., the user gives context to DQ assessment. In both articles, the same need arises. However, in \citePS{PS35} the elements that make up the context are explicitly mentioned, while in \citePS{PS23} they are never named. This means that, as in many cases the context is not defined, it must be deduced. This implies determining which are the components that make up the context. 

On the other hand, the representation of the context is linked to its components. This is so, because representing the context means representing each of its components. For this, we identify different types of representations. For example, in \citePS{PS20} data filtering needs and DQ requirements are the components of the context. In turn, these components are represented through semantic and syntactic rules. Therefore, in this section we will address the level of formalization of context definition, the various components composing the context, and the representation of those components.

%All these elements that allow us to understand what the context is, will be addressed in the following subsections.
%\cpatrick{at least give here the outline of this section}

%\cpatrick{we could indicate what are the next subsections about, and maybe precise the intuition behind  context representation and components}

\subsection{Level of formalization of context definition}

\vero{Although all PS deal with the usage of context for data quality, many of them just mention its importance but do not provide a proper definition of context. In addition, when a definition is given, it is usually informal or even fuzzy.}

\begin{figure}[hbt]
\centering
\captionsetup{justification=centering}
\includegraphics[scale=0.55]{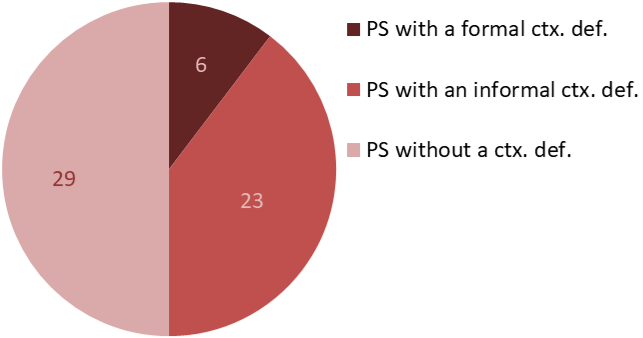}
\caption{Total of PS by level of formalization of context definition.}  
\label{fig:ctxDefType}
\end{figure}

\vero{Therefore, as a first step for understanding the underlying notion of context,} %Firstly, 
we classify the PS according to \vero{existence of a proper definition of context and its level of formalization, in three categories:} 
%three possible levels of formalization of the context definitions: 
(i) \textit{formal \vero{definition}}, \vero{when} %if 
the context is defined formally; 
(ii) \textit{\vero{informal definition}}, %no formal, 
when the context is defined, but not formally, \vero{for example,} in natural language; 
(iii) \textit{\vero{no definition}}, %implicit, 
when the context is not defined (not even in natural language), but is used implicitly. The latter occurs\vero{, for example,} when the authors present the importance of data context, but they do not define what the context is. \vero{Quantitatively,} 50\% of the selected PS \vero{provide no context definition,} %use the context implicitly, 
while only 10\% of the works present a formal context definition, \vero{as illustrated} in Fig. \ref{fig:ctxDefType}. % we can see the distribution of the corresponding percentages.

\vero{In next paragraphs, we report the level of formalization of context definition by research domain, highlighting domains with more formal proposals.}

\begin{figure}[hbt]
\centering
\captionsetup{justification=centering}
\includegraphics[scale=0.70]{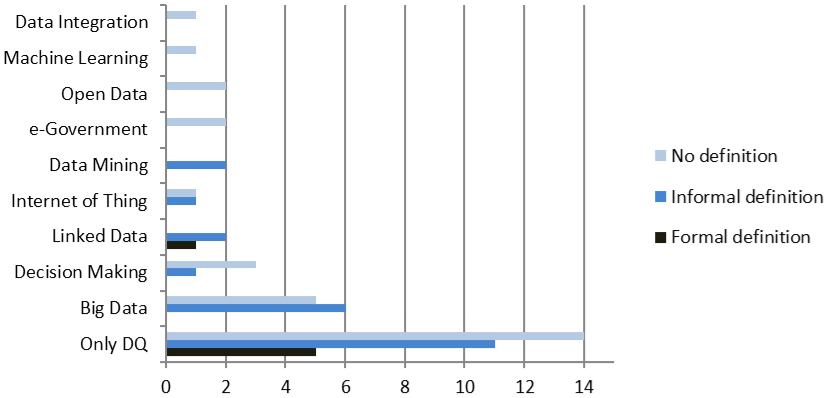}
\caption{Total of PS by research domain according to the level of formalization of context definition.}  
\label{fig:researchDomain}
\end{figure}

\paragraph{Level of formalization of the context by research domain} 
Fig. \ref{fig:researchDomain} shows the amount of PS \vero{of each level of formalization, for each research domain. A first remark is that the level of maturity of context definition is disparate, and allows a preliminary classification of research domains.}

Firstly, \textit{Only DQ} domain\vero{, the one concerning more PS (as discussed in Subsection \ref{sec:slr:results}),} is the only \vero{one presenting} the three levels of formalization. Moreover, \vero{most PS proposing formal definitions (5 PS out of 6) concern this domain; being over represented w.r.t. the overall distribution shown in Fig. \ref{fig:ctxDefType}.}
%it is the set of PS in which we find the most definitions of context, specially \cpatrick{most of them} not formals, since only 5 PS present a formal context definition. At the same time, it is also the domain that presents the most PS with an implicit context. 
These results make sense\vero{, as many theoretical proposals for DQ modeling are cross-domain.} % by the nature of the searches that were performed in the execution of the SLR. \patrick{no surprise here in the sense that this is the largest category so it somehow respects the statistics of the whole corpus}

%\vero{Interestingly, the PS proposing informal definitions concern several domains, mostly concerning \textit{Only Data/Information Quality} and \textit{Big Data Domains}.}

\flavia{On the other hand, the authors of \citePS{PS1}, a survey from 2017 focused on the evolution of DQ, highlight that organizations view Big Data, social media data, data-driven decision-making, and analytics as critical. For this, according to them, DQ has never been more important. Furthermore, they add that DQ research is reaching the threshold of significant growth from focusing on measuring and assessing DQ toward a focus on usage and context. This is reflected in several works (11 PS) that address DQ in \textit{Big Data}. These PS point out the importance of DQ, however, none of them define the context formally. Indeed, proposals concern informal or no proper definitions.} With the same needs, but to a lesser extent we have similar results, regarding the type of formalization, for the \textit{Decision Making} (4 PS) and \textit{Internet of Thing} (2 PS) domains. 

The \textit{Linked Data} (3 PS) and \textit{Data Mining} (2 PS) research domains are a special case, because although this domains presents very few PS, in all cases a context definition is given. In particular, \vero{1 PS in the} \textit{Linked Data} \vero{domain} presents a formal context definition. In this kind of domain is essential to find the best data sources, and data context plays a very important role when selecting them \citePS{PS39}. Because, it might help in interpreting the user needs. In other matters, \textit{Data Mining} domain is applied to exploit DQ. According to \citePS{PS12}, inference rules are context free, while coherent reasoning under uncertainty is sensitive to the data context. 

%\cpatrick{in those cases, can we explain why they need a definition? Are there particular reasons?}

Finally, we identify the research domains that only address context implicitly, \vero{providing no definition:} \textit{e-Government} (2 PS), \textit{Open Data} (2 PS), \textit{Machine Learning} (1 PS) and \textit{Data Integration} (1 PS). \flavia{Although there are not many PS with these characteristics, we observe that in these domains when DQ is addressed, the authors emphasize the need to identify the context of the data. However, all context definitions are given in natural language.}

%\cpatrick{interestingly, you are implicitly proposing a categorization stemming from your survey:(i) only data/information quality, (ii) big data, decision making, IOT, (iii) LOD, DM, (iv) the others. Why not presenting it this way?}

\paragraph{Summary} We highlight that we have only 6 PS that present a formal context definition, and they are proposed in the \textit{\vero{Only} DQ} and \textit{Linked Data} domains. Additionally, in a total of 10 research domains, 6 present some kind of definition for the context when managing DQ. All these results not only show the important role of context in assessing DQ \vero{in many domains}, but also the lack of formalization in the bibliography. \vero{Moreover, half of PS present no context definition, which magnifies this lack.} 

%\cvero{The next sentences seems to lead to a false conclusion: }
%Since, the implicit use of context, in almost all research domains, confirms that data are influenced by the environment in which they are used. In particular, in the \textit{Only DQ} domain, where most PS use context implicitly.

%\vero{More globally,} %Furthermore, 
Regardless of the level of formalization of the context definition, \vero{most} authors point out that the environment that influences data is defined by several components. These components are determined by how data are used, who uses data, when and where data are used, among others. \vero{They are} addressed in the next subsection.

%\cpatrick{not sure it is worth repeating what can be seen in the figure, maybe just comment salient points and write down what can be conclude}
%\cpatrick{maybe change the color of the formal bar in the barchart, and change the order of the domain axis}

\subsection{Context Components}
\label{sec:results:ctxComp}

The state of the art revealed that although there are general operational definitions of context and context-aware computing \cite{Dey01}, context representation is neglected in DQM. In particular, we highlight conclusions of Bertossi et al. \citePS{PS11} who report that the literature only deals with obvious contextual aspects of data, like geographic and temporal dimensions. As suggested by Bolchini et al. \cite{BolchiniCOQRST09}, other contextual aspects should be specified, e.g. users (person, application, device, etc.), presentations (system capabilities to adapt content presentation to different channels/devices), communities (set of relevant variables shared by a group of peers), and data tailoring (only selecting relevant data, functionalities and services). Preferences, documents content, DQ requirements and domain rules also emerge as important components (a preliminary review can be found in \cite{SerraM17}).

We consider that context not only fits a single perspective, but could be defined by elements taken from different perspectives (user, application domain, data tailoring, etc.). \flavia{Therefore, we reviewed the selected PS, and only 9\% do not propose any component for the context. In the rest, 91\% of PS, we identify or deduce (when the authors do not define the context, but suggest that DQ depends on certain elements.), the components of the context suggested in each proposal.} Indeed, we started by eliciting the proposed components and we group those proposing close concepts. We got to the ten categories of components listed below. For each one we highlight some of the PS that propose it:

\paragraph{DQ requirements} Many PS suggest that DQ requirements must be considered for an efficient DQ management. For instance, according to \citePS{PS55}, a DQ framework for a data integration environment needs to be capable of representing the varieties of user quality requirements (e.g. the level of precision or the rate of syntactic errors) and to provide mechanisms to detect and resolve the possible inconsistencies between them. 
    
\paragraph{Data filtering needs} According to \citePS{PS26}, these are requirements and expectations on data that are stated, generally implied or obligatory. They typically express concrete data needs for a specific task, for example, filtering data about patients with a certain health profile. 

\paragraph{System requirements} \flavia{The authors of \citePS{PS45}, focused on Big Data Systems, point out that requirements on data in this kind of systems involves several axes: data capability (network and storage requirements, e.g. system needs to support PostrgeSQL and MongoDB), data source (different characteristics of data sources, e.g. system must collect data from sensors), data transformation (data processing and analysis, e.g. system must support batch), data consumer (visualization, e.g. system must support processed results in text) and data lifecycle (data lifecycle management functionality, e.g. system must support DQ).}
    
\paragraph{Business rules} In \citePS{PS3} business rules are simply constraints defined over data. \flavia{While in \citePS{PS26} they are policies that govern business actions that result in constraints on data relationships and values.} They typically express conditions that data must satisfy in order to ensure the consistency of the dataset. Contrarily to data filtering needs, in general, business rules are independent of the task on hand. 
    
\paragraph{Application domain} The authors of \citePS{PS43} mention the importance of developing and tailoring quality checks to extend the effectiveness of the DQ metrics in detecting ``dirty data", and contextualizing domain characteristics. Many PS indicate that data are conditioned by the application domain. For instance, many works addressing DQ assessment, consider particular aspects of their application domains for the definition of DQ metrics. In particular, in \citePS{PS42} the authors highlight that it is possible to define general metrics, but these are often not sufficient to determine DQ problems specific to a given domain. For this reason, they suggest defining domain-specific DQ metrics.

\paragraph{Task at hand} The task performed by the user plays an important role when defining the context. The proposal in \citePS{PS18} indicates that DQ management for Big Data should prioritize those DQ dimensions really addressing the DQ requirements for the task at hand. In particular, Wang and Strong in \cite{wang-strong} underline that DQ must be considered within the context of the task at hand.
    
\paragraph{User characteristics} In most PS DQ depends on the user. They suggest several characteristics of the user that provide the context. Among them, the user profile implies general aspects of the user, such as his geographical location, language, etc. User preferences are also related to what the user likes. The proposal in \citePS{PS44} shows the relationships among perceived information quality (among others), and the perception, satisfaction, trust and demographic characteristics of the users (such as identification, gender, age, education, internet experience, etc.), in e-government environment.
    
\paragraph{Metadata} In this category, the PS consider metadata to determine data context. In \citePS{PS48}, the connection between metadata and DQ problems is investigated. Metadata, such as count of rows, count of nulls, count of values, and count of value pattern are used to generate DQ rules. The authors of this work also mention that for DQ management it is necessary to categorize metadata for improving DQ.
    
\paragraph{DQ values} This category suggest to use DQ values to give context to other DQ measures. Although these are also metadata, we consider important to have a category for them, since they are a special type of metadata. For instance, the authors in \citePS{PS22} investigate how metrics for the DQ dimensions completeness, validity, and currency can be aggregated to derive an indicator for accuracy. They highlight that although it is well known that there are interdependencies between particular DQ dimensions, their measurement are usually discussed independently from each other.

\paragraph{Other data} In this case, the quality of a dataset is evaluated based on other data that are not the contextualized data. For instance, in a relational database, data from one table could give context to other tables. Examples of this case can be seen in \citePS{PS11,PS19}.

\begin{figure}[hbt]
\centering
\captionsetup{justification=centering}
\includegraphics[scale=0.70]{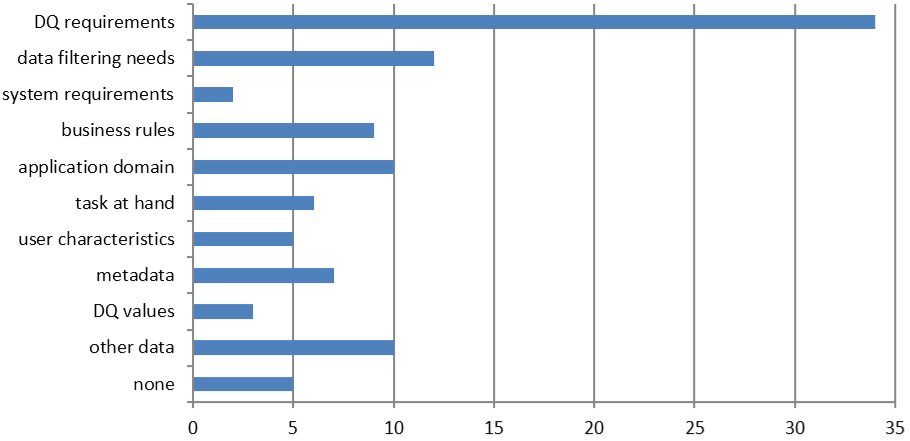}
\caption{Total of PS by context components.}  
\label{fig:totalCtxComp}
\end{figure}

In Figure \ref{fig:totalCtxComp}, we present the amount of PS that propose each of the context components. Some PS propose several of these components. \flavia{According to the results, data are mostly influenced by \textit{DQ requirements}, \textit{data filtering needs}, the \textit{application domain} and \textit{other data}, the latter are not the contextualized data. To a lesser extent, data are also influenced by \textit{business rules}. This makes sense since business rules are strongly tied to the application domain.}

In other matters, there are 5 PS where the importance of considering the context for managing DQ is highlighted, but they do not mention which are the components of such context. In the Fig. \ref{fig:totalCtxComp} these PS are classified as "none". 

\flavia{So far we have seen that there is not a single context, since it depends on the components that make it up. The components of the context vary according to the elements that have the greatest influence on the data. For instance, sometimes the context is only conditioned by the characteristics of the user, since the data depends on the geographical location, age, expertise, etc. of the user. On other occasions, the important thing is the application domain of the data, regardless of the user who uses such data. Therefore, to identify the components of the data context, it is first necessary to identify the elements that will condition the use of the data.}

\flavia{On the other hand, these elements could vary throughout the entire data life cycle, in particular through the different stages of the DQM process. For this reason, below we identify at which stages of the DQ process each of the proposed context components appears.}

%
%  \cpatrick{what is the take away message up to this point?}
%In the description of each of the suggested context components, we reference some particular PS, however, Table \ref{tab:ctxCompxDefType} shows the proposed context components by each PS. On the other hand, we are also interested in analyzing new classifications for the context components proposed. First, in Table \ref{tab:ctxCompxDefType} we also analyze context components by the level of formalization of the context definition. Secondly, we want to identify which are the context components that participate in each of the stages of the process of DQ management. These classifications are described below. 
%\cpatrick{this paragraph is not clear to me} \cvero{neither to me}

\begin{figure}[bt]
\centering
\captionsetup{justification=centering} 
\includegraphics[scale=0.50]{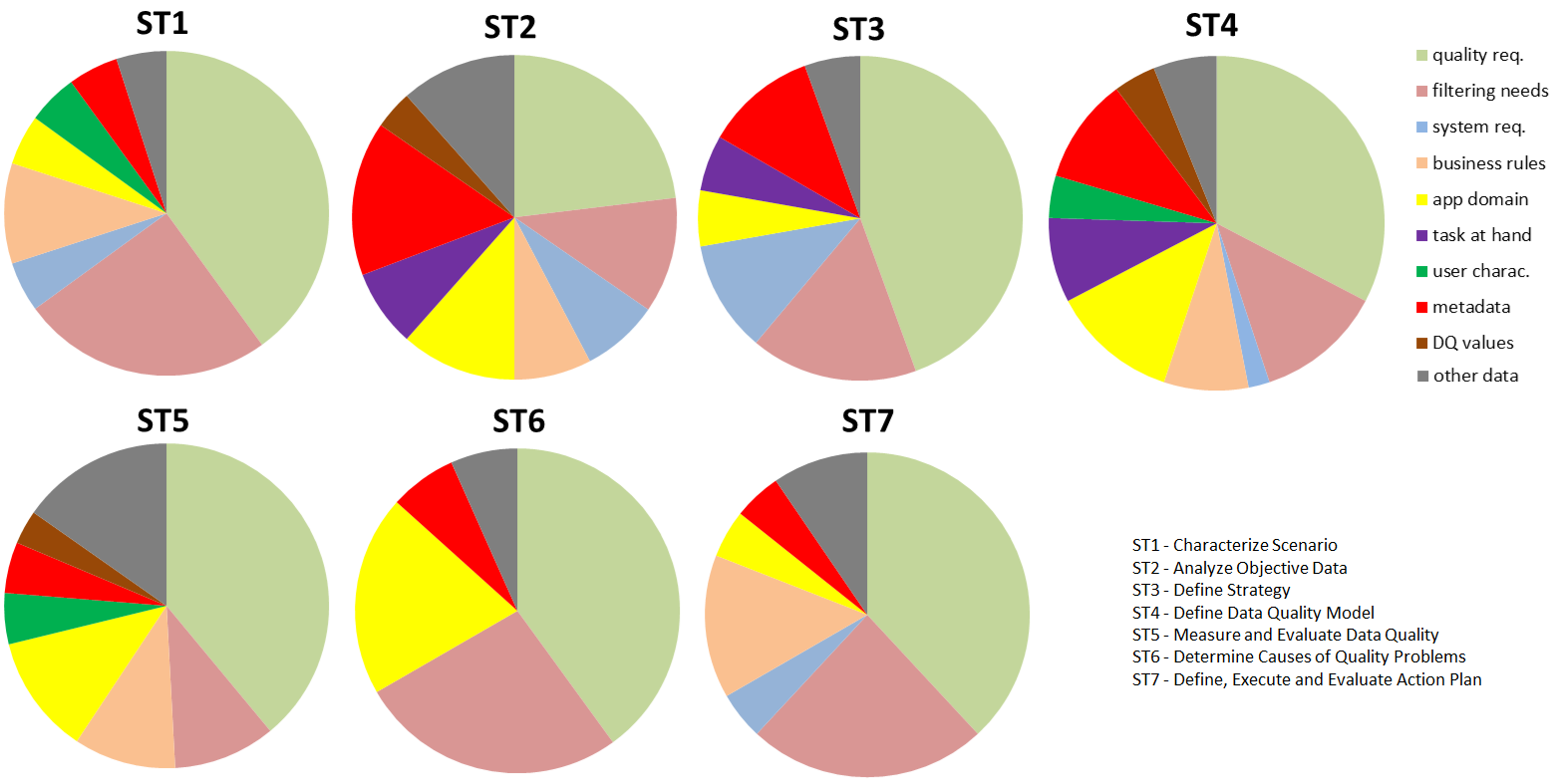}
\caption{Context components at each stage of the DQ process.}  
\label{fig:ctxCompXStages}
\end{figure}

\paragraph{Context components by DQ process stages} We \vero{now investigate} which context components are considered at each DQM process stage, in Table \ref{tab:ctxCompXstage} \flavia{we classify each PS according to the context components propose. Besides, we quantify the results and summarize them in Figure \ref{fig:ctxCompXStages}, where we show the context components that participate at each of the stages.} 

Based on this classification, it appears to make sense that some context components are more important than others at certain stages of the DQM process. \flavia{In fact, there are components that are not taken into account at some stages. For instance, DQ requirements, data filtering needs, application domain, metadata, and other data are suggested as context components at all stages of the DQM process. However, this does not happen for all the context components analyzed. Next, we will analyze the suggested components at each of the stages of the DQM process.} %In Figure \ref{fig:porcCtxComp-Stages}, we show these results quantitatively, since for each context component we show the percentage of PS that consider it at each stage of the DQM process.

%Firstly, for the stages ST1 \vero{(scenario characterization)} and ST2 \vero{(}objective data analysis\vero{),} user and other data are the favorites context components. The same happens in stage ST3 \vero{(define strategy)}, when defining a prioritization strategy for DQ requirements. 

%In the case of the stages ST4 and ST5, DQ model definition and DQ measurement and evaluation respectively, most of PS address domain, user, other data and data/quality requirements as the main context components. On the other hand, in the stage ST6, where the causes of DQ problems are addressed, fewer context components are considered. Anyway, user is the most proposed component. In the final stage ST7, where a DQ improvement action plan is defined, executed and evaluated; user, other data and guidelines are the most relevant context components.

%Finally, we have 1 PS that proposes user and other data as context components, but it is not associated with any DQM process stage. In turn, we identify 7 PS that address some of the DQ process stages, but their authors do not define which are the context components that participate in such stages. 

\paragraph{ST1 - Scenario characterization} \flavia{At this stage, the task at hand and DQ values are not taken into account as context component in any of the PS. For the former it is not an expected result, since the task at hand are part of the work scenario. For the latter, the result makes sense, because at this stage DQ values are not yet known. On the other hand, only 1 PS takes the application domain into account for the work scenario characterization. As the domain defines the work scenario, it could be a natural context component of this stage.}

\paragraph{ST2 - Objective data analysis} \flavia{In other matters, user characteristics is the only component that is not considered context component at this stage. User characteristics do not play an important role at this stage, because the data profile is analyzed as objectively as possible.}

\paragraph{ST3 - Strategy definition} \flavia{In this case, where the DQ management strategy is defined and DQ requirements are prioritized, business rules, user characteristics neither DQ values are relevant, according to the analyzed PS. However, intuitively we could believe that the definition of the DQ management strategy is strongly based on business rules and user characteristics. Besides, it seems to make sense that DQ values do not participate in the DQ management strategy. Nevertheless, the first estimation of DQ values obtained at the stage ST2, from data profile, could be useful for the prioritization of DQ requirements.}

\paragraph{ST4 - DQ model definition} \flavia{For this stage, all the PS agree that all the context components proposed influence the DQ model, where DQ dimensions and DQ metrics are defined. Based on this, this stage appears as the most context-dependent. It is in line with the concept of DQ that is defined as data that are fit for use by data consumers \cite{wang-strong}. What's more, for the area researchers, DQ is a multi-faceted concept, which is represented by DQ dimensions that address different aspects of data \citePS{PS25}, and according to Wang and Strong in \cite{wang-strong}, some of these aspects are context dependent.}

\paragraph{ST5 - DQ measurement and evaluation} \flavia{At this stage something similar to the previous stage happens. Since, except task at hand, all other components suggested for the context also influence this stage. For this reason, it is also one of the most context-dependent stages. Perhaps this result is associated with the fact that DQ community researchers have paid more attention to the stages of DQ model definition (ST4) and DQ measurement (ST5), than others stages.}

\paragraph{ST6 - Causes of DQ problems determination} \flavia{The selected PS suggest that business rules, the task at hand, user characteristics and DQ values have no influence at this stage, where the causes of DQ problems are determined. This is surprising, since the analysis of DQ problems should be conditioned to all the components that make up the context, because DQ problems could arise from several factors. In fact, DQ values should be indicators of the importance of DQ problems.}

\paragraph{ST7 - Definition, execution and evaluation of the plan} \flavia{Finally, when the action plan for DQ is defined, executed and evaluated, the task at hand, user characteristics and DQ values are not considered to form the context. In this case, it is also striking that DQ values are not taken into account in any PS, when defining an action plan. Since these indicators could condition the actions to be taken on the data, processes, users, etc. of the work scenario.}

\paragraph{Summary} \flavia{DQ requirements participate at all stages of the DQM process. Besides, according to table \ref{tab:ctxCompXstage}, they are proposed as the most relevant context components at each stage. This is a natural result, since in DQ management, DQ requirements are the starting point for measurement and evaluation. In turn, it is important not to ignore the fact that DQ requirements may change during data usage \citePS{PS26} and this can be reflected in the DQM process. Regarding data filtering needs, although they are suggested at all stages of a DQM process, where they appear with more relevance is at stages ST4 and ST5. However, at stages ST1, ST3 and ST7 they also have a very important role, since the scenario characterization, the prioritization of DQ requirements and the DQ improvement plan strongly depend on data filtering needs.}

\flavia{In other matters, business rules, user characteristics, application domain and DQ values are especially suggested at stages ST4 and ST5. The first three context components are strongly related, so it is not surprising that they appear together to form the context. Especially at the stage of DQ model definition, where DQ dimensions and DQ metrics, which can be strongly subjective \citePS{PS15}, are defined. On the other hand, the task at hand appears as context component especially at stage ST4. It is probably one of the first identified context components, since Wang \& Strong \cite{wang-strong} in 1996 emphasize that DQ must be considered within the context of the task at hand. Finally, metadata and other data condition the stages ST2, ST4 and ST5. At the stage ST2, in the objective data analysis, metadata are naturally considered.} 

\flavia{So far we have analyzed the level of formalization of the context and the components that compose it. We also saw that context can change throughout the entire DQM process. This could mean that the importance of each context component can vary at the different stages of the DQM process. On the other hand, we have not said anything about how each of the suggested components are represented. Therefore, below we will focus on analyzing the different ways of representing the context components identified. }

\begin{table}[H]
\centering
\begin{turn}{90}
\begin{tabular}{|l|c|c|c|c|c|c|c|} \hline
CTX Components & ST1 & ST2 & ST3 & ST4 & ST5 & ST6 & ST7 \\ \hline \hline

none & \citePS{PS1,PS53} & \citePS{PS1,PS53} & \citePS{PS1,PS53} & \citePS{PS1,PS53} & \citePS{PS1,PS37,PS53} & \citePS{PS1,PS29,PS30,PS53} & \citePS{PS1,PS53} \\  \hline

\multirow{6}{2cm}{DQ requirements} & \citePS{PS7,PS21,PS26,PS45} & \citePS{PS27,PS28,PS45,PS46} & \citePS{PS13,PS21,PS23,PS45} & \citePS{PS3,PS5,PS6,PS13} & \citePS{PS3,PS4,PS5,PS7} & \citePS{PS6,PS11,PS24,PS26} & \citePS{PS13,PS16,PS20,PS23} \\
 & \citePS{PS10,PS50,PS54,PS55} & \citePS{PS54,PS55} & \citePS{PS10,PS50,PS54,PS55} & \citePS{PS17,PS18,PS20,PS23} & \citePS{PS11,PS13,PS16,PS20}  & \citePS{PS34,PS56} & \citePS{PS10,PS34,PS45,PS50}  \\ 
 &  &   &  & \citePS{PS28,PS34,PS35,PS45} & \citePS{PS23,PS27,PS28,PS32} & &   \\ 
 &  &   &  & \citePS{PS10,PS55,PS56,PS58} &  \citePS{PS33,PS34,PS35,PS39} & &   \\
 &  &   &  &  &  \citePS{PS41,PS42,PS56,PS57} & &   \\ 
 &  &   &  &  &  \citePS{PS10,PS58,PS59} & &  \\ \cline{1-8} 

\multirow{2}{2cm}{data filtering needs} & \citePS{PS10,PS15,PS26,PS36} & \citePS{PS9,PS36,PS43} & \citePS{PS10,PS13,PS50} & \citePS{PS9,PS10,PS13,PS14} & \citePS{PS10,PS13,PS34,PS36} & \citePS{PS24,PS26,PS34,PS43} & \citePS{PS10,PS13,PS34,PS36} \\
 & \citePS{PS50} &   &  & \citePS{PS34,PS43}  & \citePS{PS39,PS43} & & \citePS{PS50}  \\ \cline{1-8} 
 
system requirements & \citePS{PS45} & \citePS{PS45,PS49} & \citePS{PS45,PS49} & \citePS{PS45} &  &  & \citePS{PS45} \\  \hline

\multirow{2}{2cm}{business rules} & \citePS{PS36,PS52} & \citePS{PS36,PS51} &  & \citePS{PS3,PS14,PS51,PS52} & \citePS{PS3,PS25,PS36,PS41} &  & \citePS{PS12,PS36,PS51} \\
 &  &   &  &  & \citePS{PS52,PS57} & &   \\ \cline{1-8} 

\multirow{2}{2cm}{application domain} & \citePS{PS54} & \citePS{PS9,PS43,PS54} & \citePS{PS54} & \citePS{PS5,PS6,PS9,PS34} & \citePS{PS4,PS5,PS25,PS34} & \citePS{PS6,PS34,PS43} & \citePS{PS34} \\
 &  &   &  & \citePS{PS38,PS43} &  \citePS{PS38,PS42,PS43} & &   \\ \cline{1-8} 

task at hand &   & \citePS{PS46,PS49} & \citePS{PS49}  & \citePS{PS14,PS17,PS18,PS47} &  &  &   \\  \hline

user characteristics & \citePS{PS44}  &  &  & \citePS{PS38,PS47} & \citePS{PS19,PS25,PS38} &  &   \\  \hline

\multirow{2}{2cm}{metadata} & \citePS{PS54} & \citePS{PS9,PS28,PS48,PS54} & \citePS{PS48,PS54} & \citePS{PS8,PS9,PS28,PS34} & \citePS{PS28,PS34,PS35} & \citePS{PS34} & \citePS{PS34} \\
 &  &  &  & \citePS{PS35} & & &   \\ \cline{1-8} 

DQ value &  & \citePS{PS2}  &  & \citePS{PS22,PS31} & \citePS{PS22,PS31} &  &  \\  \hline

\multirow{3}{2cm}{other data} & \citePS{PS36} & \citePS{PS2,PS28,PS36} & \citePS{PS23} & \citePS{PS23,PS28,PS35} & \citePS{PS4,PS11,PS19,PS23} & \citePS{PS11} & \citePS{PS23,PS36} \\ 
 &  &   &  &  & \citePS{PS28,PS35,PS36,PS41} & &   \\ 
 &  &   &  &  & \citePS{PS57} & &   \\ \cline{1-8} 

\end{tabular}
\end{turn}
\caption{PS by context components and DQM process stages.}
\label{tab:ctxCompXstage}
\end{table}

\subsection{Context Components Representation} 
\label{sec:results:ctxRepres}

In this section, we describe the different proposed forms to represent each component of the context, beyond the level of formalization used to define it. Of the PS that propose components to form the context, only 55\% suggest or present a representation of these components. In Fig. \ref{fig:ctxCompRep} we show the distribution of the selected PS according to the different representation proposed. On the other hand, all the suggested representations for each context component are shown in Table \ref{tab:ctxRepXctxComp}. Next, we present the context components representations and the PS that propose them:

\begin{figure}[bt]
\centering
\captionsetup{justification=centering}
\includegraphics[scale=0.70]{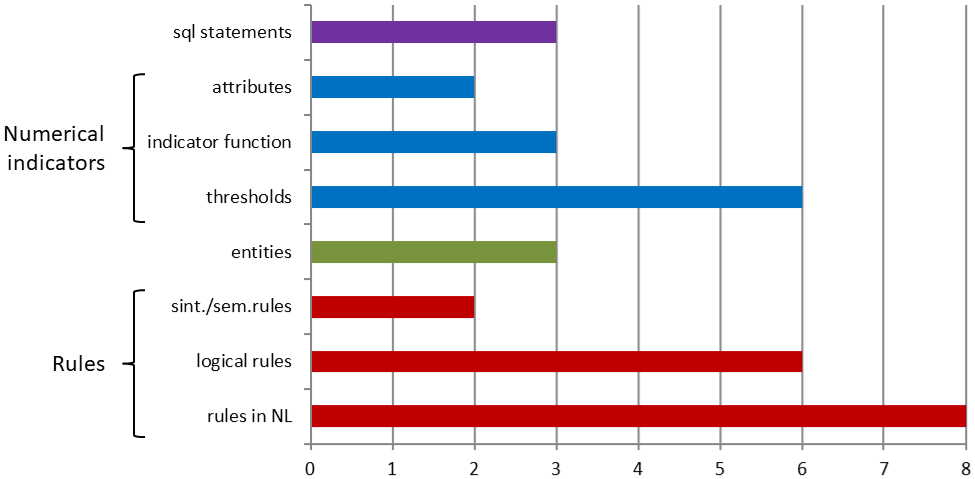}
\caption{Context component representations by PS}  
\label{fig:ctxCompRep}
\end{figure}

\paragraph{Rules in natural language} These rules are proposed to represent DQ requirements \citePS{PS10,PS20,PS23,PS27,PS42,PS45}, data filtering needs \citePS{PS10,PS36}, business rules \citePS{PS36,PS52}, system requirements \citePS{PS45} and other data \citePS{PS23,PS36}. Although the rules in natural language are the most used by PS, they are especially used to represent DQ requirements. For instance, \citePS{PS23} DQ requirements of customers, in natural language rules format, are linked to DQ dimensions, while in \citePS{PS42} they are used to develop DQ metrics. 

\paragraph{Logical rules} Also this kind of rules are widely used to represent context components. In this case, they are proposed to represent DQ requirements \citePS{PS11,PS41,PS55,PS57}, business rules \citePS{PS25,PS41,PS57}, application domain \citePS{PS25}, user characteristics \citePS{PS25} and other data \citePS{PS11,PS19,PS41,PS57}. Logical rules, although they are not the most used, they are the ones that cover the most variety of context components (in Table \ref{tab:ctxRepXctxComp}). For instance, in \citePS{PS25} the Datalog language is used to represent context components through a set of logical rules.

\paragraph{Syntactic or semantic rules} To a lesser extent, with respect to the other types of rules, syntactic or semantic rules are used for representing data filtering needs \citePS{PS43}, business rules \citePS{PS12} and application domain \citePS{PS43}. As an example, in \citePS{PS43} characteristics of the application domain and data filtering needs arising from constraints and dependencies, and they are represented through semantic and syntactic rules. In turn, in \citePS{PS12} business rules are applied as semantic rule over data items, where the items are either tuples of relational tables.
    
\paragraph{SQL statements} The proposals in \citePS{PS33,PS59} represent DQ requirements using SQL statements. In \citePS{PS33}, an extension of \citePS{PS59}, it is mentioned that SQL statements enable defining data objects using ``select'', and defining DQ requirements using ``where'' conditions. On the other hand, in \citePS{PS51} also SQL statements are used, but in this case to represent functional dependencies that arise from business rules. According to the authors, conditional functional dependencies have a syntax that can clearly represent context rules to document dependencies between attributes. This syntax also makes easier to translate rules to SQL queries that look for records that violate the rules.
    
\paragraph{Numerical indicators as thresholds} Another way to specify context components is through numerical indicators, and in this case, they are thresholds that represent DQ requirements \citePS{PS27,PS39,PS50,PS56}), application domain \citePS{PS42}, and other data \citePS{PS4}. For example, the authors in \citePS{PS39} use numerical quality indicators for a better linked data source selection. That is, user indicates thresholds for DQ values, then a ranked list of relevant sources are returned according to the quality thresholds. Quality thresholds represent DQ requirements that characterize DQ dimensions considered relevant for the use case at hand. 

\paragraph{Numerical indicators applying functions} In this case, numerical indicators are calculated by applying functions.The context components represented with this representation are metadata \citePS{PS48} and DQ values \citePS{PS22,PS31}. In \citePS{PS48} an indicator function maps the set of metadata to binary values 0 or 1, where the result 1 denotes that the respective dataset value is incorrect, otherwise it is 0, meaning ``clean data''. On the other hand, the indicator function in \citePS{PS22} is designed as a product of the results of DQ metrics for completeness, validity and currency. The indicator function in \citePS{PS31} is also designed as a product of the results of DQ metrics, but in this case DQ metrics are for completeness, readability and usefulness. In addition, in this proposal each DQ metric has an associated weight, a scalar value between 0 and 1.

\paragraph{Numerical indicators as attributes} Numerical indicators as attributes are used to represent data filtering needs \citePS{PS9}, application domain \citePS{PS9} and metadata \citePS{PS8,PS9}. In \citePS{PS8}, the authors identify attributes for information quality, and these are assigned to DQ dimensions. For instance, they associate the degree of closeness of its value v to some value v’ with accuracy. In other matters, in \citePS{PS9}, the authors suggest to ask ``how can data be characterized?'' instead of ``what causes DQ deficiencies?'', and they explain that the methodology consists of analyzing numerical indicators at the stages of data profiling, applying different tests that depend, not only on data, but also on the application domain and data filtering needs.
    
\paragraph{Entities} Context components represented using entities are data filtering needs \citePS{PS39,PS50}, application domain \citePS{PS38} and user characteristics \citePS{PS38}. The proposal in \citePS{PS38} uses a set of entities to represent the context. In addition, this PS relies on the context definition of Dey \cite{Dey01}, and although the authors present illustrative examples to show their proposal, they never include the suggested representation of context components in them. On the other hand, in \citePS{PS39,PS50}, concepts that are entities categories identified by URIs, are used for representing a single context component, data filtering needs. Notably, \citePS{PS39} is an extension of \citePS{PS50}. The proposal uses SKOS (Simple Knowledge Organization System)\footnote{https://www.w3.org/TR/skos-reference/} for defining, publishing, and sharing concepts over the Web. It is a W3C recommendation. SKOS concepts represent different data filtering needs that may be required by users to select linked data sources.

\begin{table}[H]
\centering
\begin{turn}{90}
\begin{tabular}{|l|c|c|c|c|c|c|c|c|c|c|c|c|}
\hline

\multirow{3}{*}{\textbf{Ctx Components}} & \multicolumn{8}{c|}{\textbf{Context Components Representations}} \\ \cline{2-9}

 & \multicolumn{3}{c|}{Rules} & \multicolumn{1}{c|}{Entities}  & \multicolumn{3}{c|}{Numerical indicators}  & \multicolumn{1}{c|}{Statements} \\ \cline{2-9} 
            & Nat.Lang. & Logical & Sint./Sem. &   & Thresholds & Indicator function & Attributes & SQL  \\ \hline \hline 
            
DQ requirements & X  & X &  & & X &  & & X  \\ \hline

data filtering needs & X & & X & X &   &   & X &  \\ \hline

system requirements & X & &  &  &   &   &  &  \\ \hline

business rules & X & X & X &   &  & & & X   \\ \hline

application domain & & X & X & X & X &   & X &  \\ \hline

task at hand &   &  &  &   &   &   & &   \\ \hline

user characteristics & & X &  & X & & & &   \\ \hline

metadata & & & & & & X & X &  \\ \hline

DQ values & & & & & & X & &  \\ \hline

other data & X & X & & & X & & & \\ \hline

\end{tabular}
\end{turn}
\caption{Representations suggested by the PS, for each context component.}
\label{tab:ctxRepXctxComp}
\end{table}

\paragraph{Summary} The most common ways to represent context components among the PS analyzed is using rules, especially rules in natural language and logical rules. Second are numerical quality indicators, particularly those that represent thresholds. 11 PS consider this way of representing context components, especially when it is necessary to represent metadata, and in particular applying functions for DQ values. Regarding the representation of context components through sql statements, considering the arguments of who proposes it, it seems natural this kind of representation. However, based on the bibliography consulted (only 3 PS), there is not enough evidence of its usefulness. In turn, for business rules, only SQL statements and rules are suggested for their representation, the latter in all the proposed forms. 

In the case of representation through entities, we consider that the identified cases are specific to a task, since they are used for specific needs in linked data domain. In \citePS{PS38}, such representation is suggested, but it is never used in the proposed examples. On the other hand, as mentioned in the previous section, although the task at hand is probably one of the first context components identified in the bibliography, it is the only context component for which the analyzed bibliography does not suggest any type of representation for it.

%% file: _examples.tex
\section{Description of proposals formalizing context}
\label{sec:examples} %results:ctxForm}

All selected PS underline the importance of context in DQ assessment. However, as seen in Figure \ref{fig:ctxDefType}, only 6 PS present a formal context definition. In this section, we focus on describing in detail the proposals \adri{of each}
%present in each of 
PS that gives a formal definition of the context.

\subsection{A Methodology to Evaluate Important Dimensions of Information Quality in Systems. Proposals of Todoran, Lecornu, Khenchaf \& Le Caillec 2015 \citePS{PS38}} 

In this PS, a methodology is proposed for assessing the quality of the information proposed by an information system (IS). The IS is decomposed into modules to define the quality of the information locally in each of the modules. According to the authors, local DQ assessment allows an IS analyst to check the performance of each module depending on the application context. Also, they add that data are context independent, but information is context dependent, i.e. data are \textit{put in context} and transformed into information. In this way, a simplified IS makes easier to pass from data to information. Hence, they underline that information consists of organized data having a meaning in the application context. Based on this, they use the context definition of \cite{Dey01} which says the context can be defined as any knowledge that can be used to characterize the situation of an entity. 

\begin{figure}[H]
\centering
\captionsetup{justification=centering}
\includegraphics[scale=0.5]{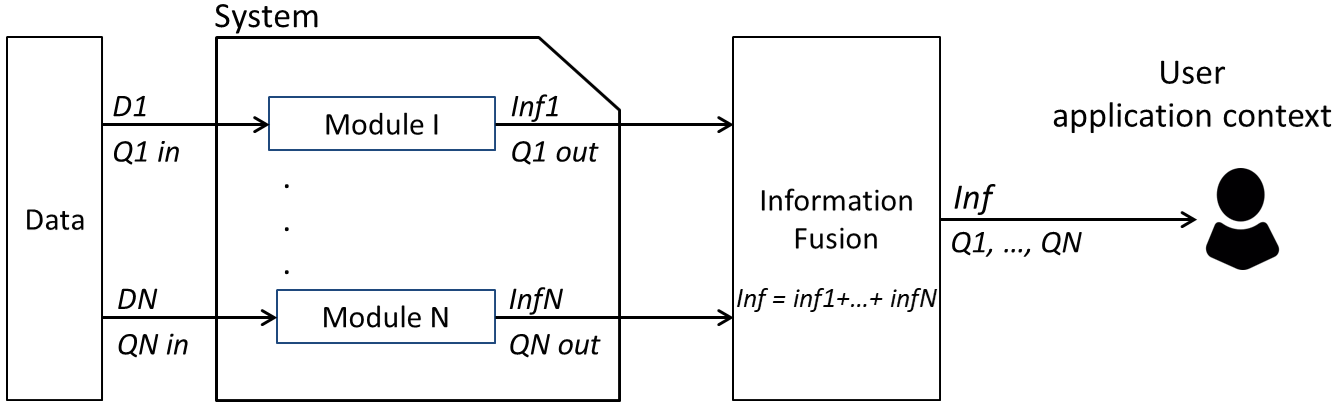}
\caption{Simplification of the proposed architecture in \citePS{PS38}.}
\label{fig:todoran_38}
\end{figure}

\paragraph{Description}
The proposal considers three major components: data, an IS and information. These components can be seen in Fig. \ref{fig:todoran_38}. The IS is decomposed into N modules. At the input of each module there is a dataset (in Fig. \ref{fig:todoran_38} \textit{D1} for module 1) that will be processed within the module to generate output information (\textit{Inf1}); and a set of DQ measures (in Fig. \ref{fig:todoran_38} \textit{Q1 in} for module 1). The input DQ measures are evaluated by different quality criteria (it could be accuracy, relevancy, completeness, currency, etc.), to generate the output DQ measures (in Fig. \ref{fig:todoran_38} \textit{Q1 out} for module 1). The information generated in each module (\textit{Inf 1}, ..., \textit{Inf N}) is merged at the stage called \textit{information fusion}. Additionally, DQ measures (\textit{Q1 out}, ..., \textit{QN out}) received at this stage are aggregated taking into account the merged information and the user application context. Finally, the DQ measures obtained (\textit{Q1, ..., QN}) and the merged information (\textit{Inf}) are delivered to the user. These DQ measures describe the merged information quality.

\paragraph{Formalization} This proposal takes the definition of context given in \cite{Dey01}, where context is \adri{defined} %defining
as any knowledge that can be used to characterize the situation of an entity (anything relevant to the interaction between a user and an application, including the user and the application). Based on this, the authors define for a given application \textit{A}, its \textit{context environment} that is defined as a \textit{set of n entities} called \textit{E1, E2, . . . En}. At the same time, each of these entities is characterized by a \textit{context domain} called \textit{dom(Ei)} that is an infinitely countable set of values.

\paragraph{Summary} 
The authors underline that for turning data into information, some knowledge is needed, and it is given by the context. The context definition is taken from the bibliography \cite{Dey01}, and although they propose a formal representation of the data/information quality, considering the context for applying DQ dimensions (called by the authors, quality criteria), they do not apply in their examples the proposed formalization. %for the context definition used. %Since (ADRIANA)
In the examples, they only use the context in an implicit way.

\subsection{Exploiting Context and Quality for Linked Data Source Selection. Proposals of Catania, Guerrini \& Yaman 2019 \citePS{PS39}} 

This work is an extension of \citePS{PS50}, and addresses the problem of source selection for Linked Data, relying on context and user DQ requirements. \adri{The authors claim that due to the semantic heterogeneity of the web of data, it is not always easy to assess relevancy of data sources or quality of their data. They consider that context information can help in interpreting users needs. In their proposal,} the sources are distributed in a data summaries structure that supports context and quality aware source selection. %The authors consider that context information can help in interpreting users needs. 
%Because due to the semantic heterogeneity of the web of data, it is not always easy to assess relevancy of data sources or the quality of the data that they contain.

%This structure is called \textit{QTree} and is an approach for determining which sources may contribute answers to a query in distributed live query systems.
\begin{figure}[H]
\centering
\captionsetup{justification=centering}
\includegraphics[scale=0.4]{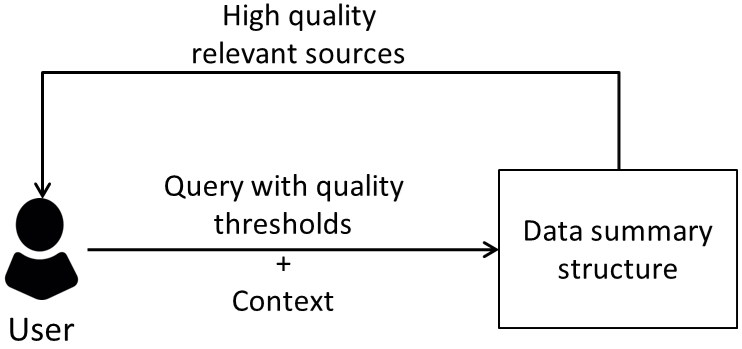}
\caption{Simplification of the proposal in \citePS{PS39}.}
\label{fig:catania_39}
\end{figure}

\paragraph{Description}
In Fig. \ref{fig:catania_39} we present a simplification of the proposal of this PS, focusing mainly on the elements that interest us for our research. We can see, the user submits a SPARQL query to the system, and this is accompanied by a set of quality thresholds \textit{<v1,..., vn>} and a reference context \textit{c}. Each \textit{vi} is a value between 0 and 1, and it specifies that the user is interested in values greater than \textit{vi}. According to the authors, quality thresholds characterize DQ dimensions considered relevant for the use case at hand. %On the other hand, an example reference context could be \textit{Tourist\_attractions\_in\_Europe}.

During a search with respect to the query, it first look for sources that satisfy the quality thresholds specified by the user through \textit{<v1,..., vn>} (i.e. user DQ requirements). If sources are found, they are ranked according to the \textit{context distance}. The latter is the distance between the contexts associated with the selected source and the reference context \textit{c} of the user query. Finally, the user receives the most relevant sources according to his/her reference context, whose data fits \adri{his/her DQ requirements.}
%the DQ requirements of the user.

\paragraph{Formalization} In this proposal the context is modelled through SKOS\footnote{Simple Knowledge Organization System: a W3C recommendation for developing, sharing and linking knowledge organization systems via the Web; https://www.w3.org/TR/skos-reference/} concepts, and these concepts are identified through URI references. According to the authors, SKOS is a W3C recommendation for defining, publishing, and sharing concepts over the Web. Besides, they add that although SKOS does not provide as powerful semantic and reasoning capabilities as ontologies, concepts enhance the potential of search capabilities. %Moreover, 
They also mention that SKOS provides mappings between concepts from different concept schemes and in diverse data sources, as well as a hierarchical organization of concepts.

\paragraph{Summary} 
In this PS, the context of the use case at hand and DQ requirements of the user are considered for the selection of data sources for Linked Data. Contexts are modelled through SKOS concepts and user DQ requirements, expressed as quality thresholds, \adri{which} characterize DQ dimensions of interest to the user. Data sources that satisfy the quality thresholds and that most closely approximate to the %concept that represents the 
reference data context, will be considered the most relevant \adri{ones} and with the highest quality.

\subsection{Rule-Based Multidimensional Data Quality Assessment Using Contexts. Proposal of Marotta \& Vaisman 2016 \citePS{PS19}}

This proposal uses logic rules to assess the quality of measures in a Data Warehouse (DW), taking into account the context in which these DW measures are considered. 

\begin{figure}[H]
\centering
\captionsetup{justification=centering}
\includegraphics[scale=0.4]{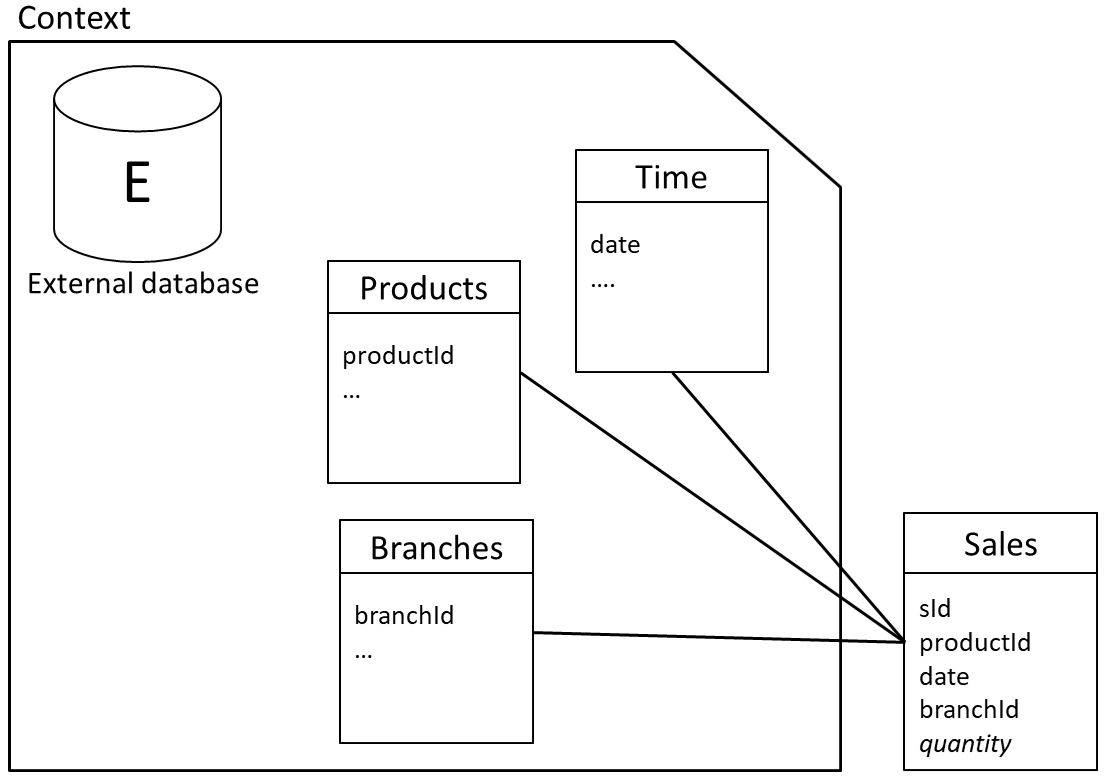}
\caption{Representation of the proposed context in \citePS{PS19}.}
\label{fig:vaisman_19}
\end{figure}

A DW is a database supporting decision making \adri{and whose data is represented and manipulated through the Multidimensional Model (MD)}. Real-world facts (e.g. \textit{sales}) are represented by a set of dimensions (describing different analysis axes), and a set of measures (numerical values associated to facts, as \textit{quantity sold}). Data are organized in star schemes, containing a central table called fact table (\textit{Sales} in Fig. \ref{fig:vaisman_19}), storing facts (and their measures, e.g. \textit{quantity} in Fig. \ref{fig:vaisman_19}) and dimension tables describing dimensions (\textit{Branches} of a supermarket, \textit{Products}, and \textit{Time} in Fig. \ref{fig:vaisman_19}).

%The DWs are based on a multidimensional model that consists of dimensions and facts. DW dimensions represent the different perspectives used to analyze the data. This information is stored in the dimensional tables. On the other hand, the facts are associated with numerical values, called measures. DW measures allow the quantitative evaluation of different aspects of the analysis problem. This information is stored in the fact tables.
%A multidimensional data model consists of facts and dimensions, where facts represent a combination of dimensions EXPLICAR EN POCAS PALABRAS

\paragraph{Description}
As can be seen in Fig. \ref{fig:vaisman_19}, the components of the context proposed in this PS are the DW dimensions (\textit{Branches, Products, and Time}) and data external to the DW (database \textit{E}). The authors propose to use logic rules to assess the quality of the DW measure (\textit{quantity}), taking into account the context. It is worth noting that in a more complex situation, the authors also include, in the context, final users of the DW system. 

DQ metrics usually represent aggregate values (e.g. accuracy is measured as the ratio of measure values that are accurate) or Boolean values (e.g. a specific value is accurate or not). However, in this proposal, a DQ metric for accuracy dimension has the following form: \textit{A DW measure is likely to be inaccurate if the quantity sold of a product in the chocolate family, in January in Uruguay is greater than 50}. Then, the rule that represents this DQ metric is executed to select the facts, from the fact table, that satisfy this DQ rule.

%\cpatrick{this metric here is, again, an intentional definition. but is this really a DQ metric? I would expect some form of aggregation, something like: accuracy is measured as the ratio of measure values that are accurate. Or maybe just a boolean indicating if a specific value is accurate or not?}
%\cpatrick{in conclusion, context corresponds to intentional definition, ie a query. now it should be clear why this query needs to be expressed in datalog and not in eg relational algebra or even conjunctive algebra}

%The proposal is conceived for star schemas. The fact table is described through the combination of predicates, one that represents each fact as an abstract entity and other that attaches the abstract facts to the DW dimensions instances. The context is defined as rules, and these rules give context to the facts measures. DQ metrics are defined also as rules, and these are used for detecting quality problems in the facts measures. Context rules incorporate DW dimensions data, called internal data, as well as external data also given by predicates.

\paragraph{Formalization} The formalization of context and DQ metrics are given as logic rules. Also, the MD data model is represented as a collection of axioms and logic rules, based on the idea proposed by \cite{Minuto02}. The authors mention that they have limited themselves to non-recursive Datalog, and %inclusive 
they don't even use negation. They underline that they could have also used it to express more complex queries, but they consider that it was not the objective of the proposal, since the main objective was to show the applicability of the approach. Additionally, the researchers consider that even the most basic Datalog version would allow them expressing context for multidimensional data, not only using DW dimensions but external data too.

\paragraph{Summary} 
According to the authors, \citePS{PS19} proposes a multidimensional model which uses Datalog rules to represent facts, dimensions, and aggregate data. In addition, they apply DQ metrics, depicted by DQ rules that are defined taking into account the context of the DW measures, for obtaining a set of facts that satisfy such rules.

\subsection{Data Quality Is Context Dependent. Bertossi, Rizzolo \& Jiang 2011 \citePS{PS11}} 

In this work, authors propose DQ assessment and DQ query answering as context-dependent activities. They mention that DQ is usually related to the discrepancy between the stored values and the real values that were supposed or expected to be stored. However, they focus on another type of semantic discrepancy. Specifically, this proposal focuses on DQ problems caused by semantic discrepancy that occurs when senses or meanings attributed by the different agents to the actual values in the database do not coincide. %On the other hand, 
The context formalization is given as a data and metadata integrated system, called contextual system.

\begin{figure}[H]
\centering
\captionsetup{justification=centering}
\includegraphics[scale=0.4]{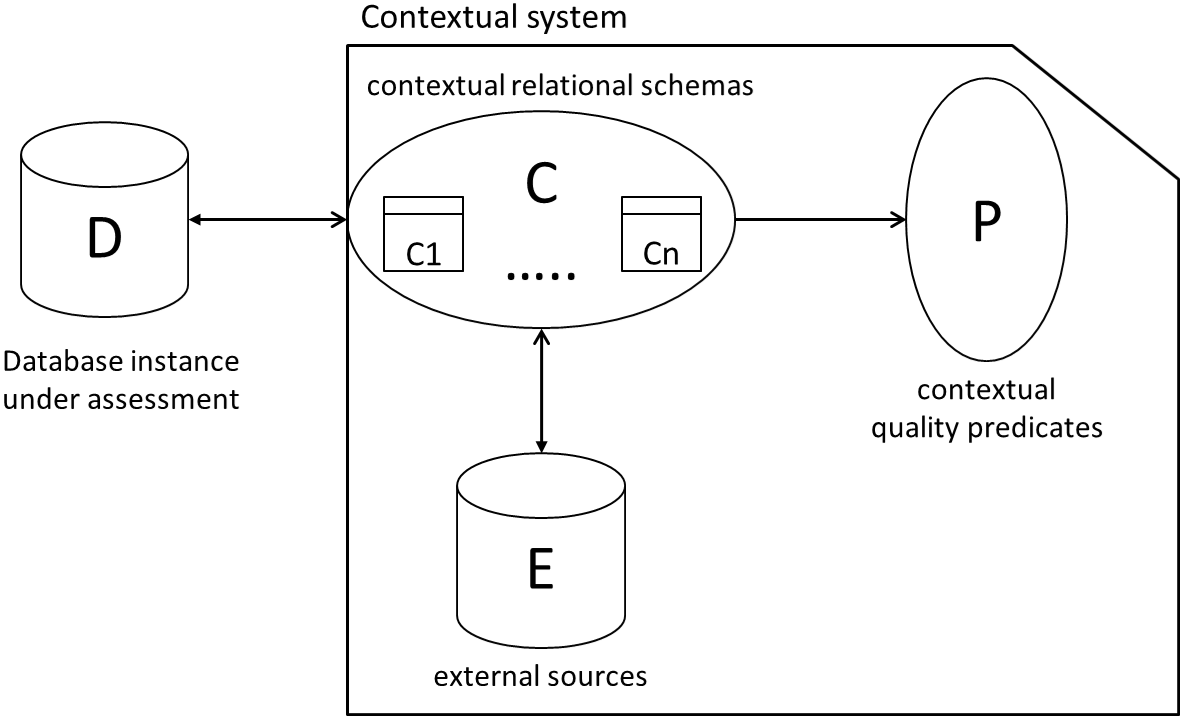}
\caption{Simplification of the general framework for contextual DQ proposed in \citePS{PS11}.}
\label{fig:bertossi_11}
\end{figure}

\paragraph{Description}
The authors propose a contextual system for modeling the context of an instance \textit{D} of a relational schema, i.e. \textit{D} is under quality assessment with \adri{respect} %reference 
to such contextual system. Fig. \ref{fig:bertossi_11} presents a simplification of the framework for contextual DQ. Thus, the contextual system consists of a set of contextual relational schemas \textit{C} = \textit{C\textsubscript{1},..., C\textsubscript{n}}; a set of \textit{contextual quality predicates} which are defined over \textit{C}, and in a more complex scenario, also a set of schemas from \textit{external sources}. Furthermore, DQ requirements represent the conditions that data must verify to satisfy the information needs of the users. Contextual quality predicates capture these DQ requirements. 

\paragraph{Formalization} The proposal considers a relational schema S, with relational predicates \textit{R\textsubscript{1},..., R\textsubscript{n}} $\in$ \textit{S}, of first-order logic. The authors consider an instance \textit{D} of \textit{S}, and the database instances are conceived as finite sets of ground atoms. For each \textit{R} $\in$ \textit{S}, instances \textit{D} are those under quality assessment wrt the contextual system. Besides, the contextual relational schema \textit{C} may include a set of predicates \textit{C\textsubscript{1},..., C\textsubscript{n}}. Each contextual quality predicate is defined as a conjunctive view in terms of elements of \textit{C}, and each database predicate \textit{R} $\in$ \textit{S} is also defined as a conjunctive view. The authors mention that they consider only monotone queries and views (e.g. conjunctive queries) that they write in non-recursive Datalog with built-ins.

%The authors also consider an instance \textit{D} of \textit{S}, and according to them, if database instances are conceived as finite sets of ground atoms, for each \textit{R} $\in$ \textit{S}, instances \textit{D} are those under quality assessment wrt the contextual system. 
%Regarding the latter, in its simplest form, it consists of a contextual relational schema \textit{C} that may include a set of predicates \textit{C\textsubscript{1},..., C\textsubscript{n}}. In turn, each contextual quality predicates is defined as a conjunctive view in terms of elements of \textit{C}, and each database predicate \textit{R} $\in$ \textit{S} is also defined as a conjunctive view. The authors mention that they consider only monotone queries and views (e.g. conjunctive queries) that they write in non-recursive Datalog with built-ins.

\paragraph{Summary}
Authors propose a framework for the assessment of a database instance in terms of quality properties. These properties are alternative instances that are obtained by interaction with additional contextual data or metadata. In turn, contextual quality predicates capture DQ requirements. On the other hand, although the proposal is motivated for supporting DQ assessment, only deal with the selection of appropriate data for query answering. Authors do not address DQ assessment issues, nor specific DQ dimensions or DQ metrics. In fact, the authors mention that the contextual schema and data are not used to enforce quality of a given instance, but rather the quality of the data in the instance is evaluated. Finally, the contextual schema characterizes the quality answers to queries.

\subsection{Ontological Multidimensional Data Models and Contextual Data Quality. Bertossi \& Milani 2018 \citePS{PS41}, extension of Milani, Bertossi \& Ariyan 2014 \citePS{PS57}} 

The framework presented in \citePS{PS11} (and previously described) is extended in the works \citePS{PS57, PS41}. Specifically, these extensions present a formal model of context for context-based DQ assessment and quality data extraction. For that, the authors propose ontological contexts including multidimensional (MD) data models in them. In this work, DQ refers to the degree to which data fits a form of usage, and it relates DQ concerns to the production and the use of data.

\begin{figure}[H]
\centering
\captionsetup{justification=centering}
\includegraphics[scale=0.4]{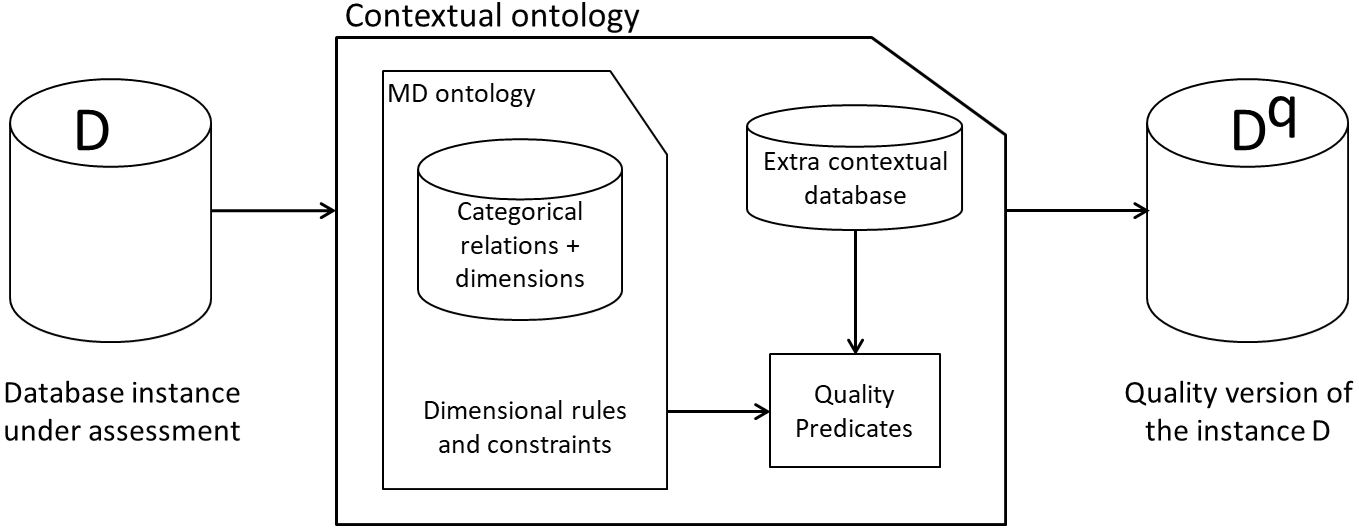}
\caption{Simplification of the contextual ontology proposed in \citePS{PS41}.}
\label{fig:bertossi_41}
\end{figure}

\paragraph{Description}
The authors underline that the instance of a given database does not have all the necessary elements to assess DQ or to extract data of quality. For this reason, they consider that the context provides additional information about the origin and intended use of data. Therefore, the authors use MD ontologies for representing the context, and they extended these ontologies with rules for DQ specification. A ontology contains definitions of quality predicates that are used to produce quality versions of the original tables. As seen in Fig. \ref{fig:bertossi_41}, data in the given database instance \textit{D} are \textit{put in context} to get a quality version of \textit{D}. This means that data obtained from \textit{D} are processed through the contextual ontology to obtain quality data.

According to the researchers, the contextual ontology contains a multidimensional core ontology. i.e. the ontological multidimensional data model represented by \textit{categorical relations and dimensions}, and eventually \textit{dimensional rules and constraints}. Additionally, a quality-oriented sub-ontology that contains \textit{quality predicates} (rules and/or constraints ), and possible \textit{extra contextual database}. The latter could be used at the contextual level in combination with data associated to the multidimensional ontology. Fig. \ref{fig:bertossi_41} shows a simplification of the proposed contextual ontology, where it is presented how these elements are related.

\begin{figure}[H]
\centering
\captionsetup{justification=centering}
\includegraphics[scale=0.4]{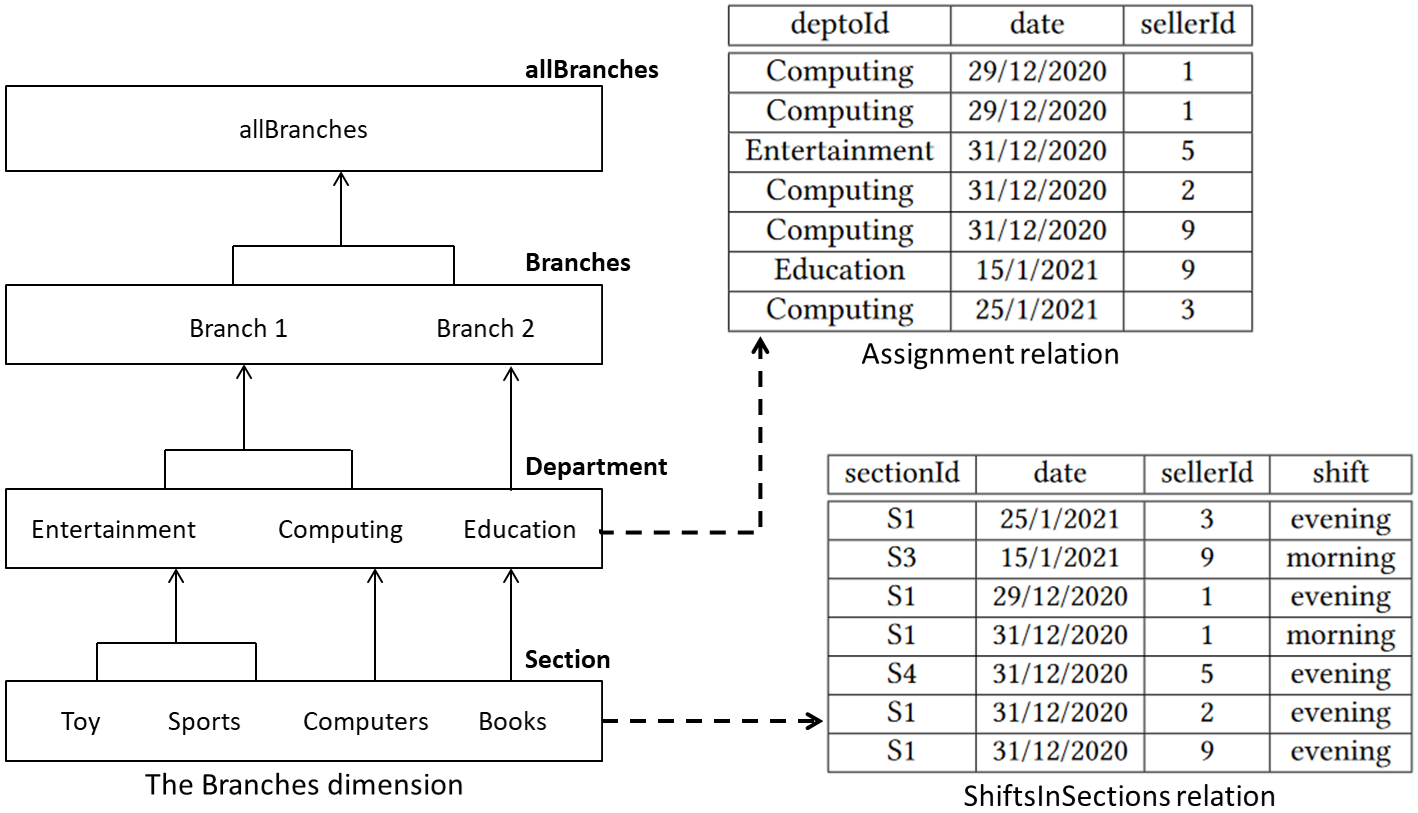}
\caption{Dimensional data example with categorical relations according to \citePS{PS41}.}
\label{fig:dimensionRelations}
\end{figure}

On the other hand, in Fig. \ref{fig:dimensionRelations} we show how are proposed the dimensional data with categorical relations. Each level of the hierarchical dimension corresponds to a relations that contains more information for each data of the level. Then, data in D is processed based on contextual information, called guidelines. In the example of Fig. \ref{fig:dimensionRelations} a guideline could be \textit{"Sellers, in the computing department, update customer data at the time of the sale"}, and it will condition the navigation in the dimensional hierarchy (selecting the level and the level data, according to the contextual information), and the selection of the data in the categorical relations. Finally, quality predicates, and eventually extra contextual data, will be taken into account to obtain a quality version of D.

\paragraph{Formalization} The authors extend the formal context model presents in \citePS{PS11} for context-based DQ assessment, DQ extraction, and data cleaning on a relational database. In this case, they are focus on an ontological context. Regarding the representation of the contextual ontology, they consider that it must be written in a logical language, and propose Datalog$\pm$ for this task. Also a declarative query language for relational databases is taken into account. According to the authors, it provides declarative extensions of relational data through expressive rules and semantic constraints.

\flavia{The work in \cite{HMmodel} is applied in this proposal, which uses Datalog$\pm$ to represent multidimensional data model. Bertossi et al. mention that certain classes of Datalog$\pm$ programs have non-trivial expressive power and good computational properties. The researchers add that some of these programs allow them to represent a logic-based, relational reconstruction and extension of the proposal in \cite{HMmodel}.}

\paragraph{Summary}
This work proposes ontological contexts embedding multidimensional data models in them. The notion of quality presented in this proposal has to do with the DQ dimension called \textit{Relevancy} \cite{wang-strong}. This PS highlights that this contextual approach can be used to address inconsistency, redundancy and incompleteness of data. Nevertheless, authors do not delve into these DQ dimensions and do not define DQ metrics for such dimensions, since it is specifically focused on query answering. In fact, the authors mention that independently from the DQ dimension, they can consider DQ assessment and data cleaning as context-dependent activities.

\subsection{General results} 
As a summary, we want to highlight that in two cases \citePS{PS38, PS41} appears the need to \textit{put data in context}. In the former to obtain information and in the latter to obtain a quality database. Although all works focus on data context, such data are considered at different levels of granularity: a single value, a relation, a database, etc. For instance, in \citePS{PS19} dimensions of a Data Warehouse (DW) and external data to the DW give context to DW measures. While, in \citePS{PS11} data in relations, DQ requirements and external data sources give context to other relations. 
On the other hand, proposals in \citePS{PS41, PS57} model and represent a multidimensional contextual ontology. The context, in this cases, is defined by hierarchical data, DQ requirements and external data sources, and all these give context to relational databases. The authors in \citePS{PS39} propose a framework where the context (represented by SKOS concepts), and DQ requirements of users (expressed as quality thresholds), are using for selecting Linked Data sources. 

In turn, \citePS{PS38} presents an information quality methodology that considers the context definition given in \cite{Dey01}. This context definition is represented through a context environment (a set of entities), and context domains (it defines the domain of each entity). In this case, researchers assume that although the context definition given is too general for DQ domain, it allows them to enumerate the context elements for a given application scenario. Next, in Table \ref{tab:tableFormalProposal} we summarize this information.

\begin{table}[H]
\begin{center}
\begin{tabular}{ | m{0.5cm} | m{2.8cm}  | m{3.2cm} | m{3.5cm} | }
\hline 
\textbf{PS} & \textbf{CTX components}  & \textbf{CTX representation}   & \textbf{Contextualized object} \\ \hline

\citePS{PS11} & DQ req., other data & relational schema & relational DB  \\ \hline

\citePS{PS57} & DQ req., business rules, other data & ontology & relational DB  \\ \hline

\citePS{PS41} & DQ req., business rules, other data & ontology & relational DB \\ \hline

\citePS{PS38} & user, app. domain & set of entities & information \\ \hline

\citePS{PS39} & DQ req., data fil. needs & SKOS concept & data \\ \hline

\citePS{PS19} & other data, user & logic rules & DW measures \\ \hline

\end{tabular}
\caption{Context characteristics in PS with context formal definitions.}
\label{tab:tableFormalProposal}
\end{center}
\end{table} 

Secondly, we are interested in identifying the most important DQ concepts addressed in these PS. For instance, \textit{DQ requirements} are addressed by all PS except in \citePS{PS38}. \citePS{PS19,PS38,PS39} consider contextual \textit{DQ dimensions}, while in \citePS{PS41} DQ dimensions are non-contextual and in \citePS{PS11,PS57} DQ dimensions are not mentioned. Regarding \textit{DQ metrics}, they appear in \citePS{PS19,PS38,PS39}, and in all of them they are contextual, i.e. their definition includes context components or they are influenced by the context. In the case of \textit{DQ tasks}, cleaning \citePS{PS11,PS41,PS57}, measurement \citePS{PS19} and assessment \citePS{PS38,PS39} are the only tasks tackled in these PS. We summarize this information in Table \ref{tab:tableFormalProposal_DQ}.

\begin{table}[H]
\begin{center}
\begin{tabular}{ | m{0.5cm} | m{1.2cm} | m{2.5cm}  | m{1.8cm}  | m{2cm} | }
\hline 
\textbf{PS} & \textbf{DQ req.} & \textbf{DQ dimensions} & \textbf{DQ metrics} & \textbf{DQ tasks} \\ \hline

\citePS{PS11} & \centering \checkmark & none & none & cleaning  \\ \hline
\citePS{PS57} & \centering \checkmark & none & none & cleaning\\ \hline
\citePS{PS41} & \centering \checkmark & non-contextual & none & cleaning  \\ \hline
\citePS{PS38} & \centering $\times$ & contextual & contextual & assessment \\ \hline
\citePS{PS39} & \centering \checkmark & contextual & contextual & assessment  \\ \hline
\citePS{PS19} & \centering \checkmark & contextual & contextual & measurement \\ \hline
\end{tabular}
\caption{DQ concepts in PS that present a context formal definition.}
\label{tab:tableFormalProposal_DQ}
\end{center}
\end{table} 

Finally, we present the most important characteristics of these PS. Regarding the research domain, \citePS{PS19,PS39} address context definitions for Data Warehouse Systems and Linked Data Source Selection respectively. On the other hand, \citePS{PS38,PS11,PS41,PS57} are specifically focused on DQ, the last three proposals tackle cleaning and DQ query answering. Furthermore, the type of work most approached is the definition of a model \citePS{PS11,PS19,PS39,PS41,PS57}. In the case of \citePS{PS41,PS57}, they also present a contextual ontology, while \citePS{PS19,PS39} also pose a framework and \citePS{PS38} only presents a DQ methodology. According to the venues quality, \citePS{PS38,PS39,PS41} are ranked with B and \citePS{PS11,PS19,PS57} are not ranked. Regarding to the publication date, the oldest PS corresponds to the year 2010 and the newest to 2019. This information is shown in Table \ref{tab:tableFormalProposal_I}.

\begin{table}[H]
\begin{center}
\begin{tabular}{ | m{0.5cm} | m{2.5cm} | m{2.5cm}  | m{2.5cm} | m{1.5cm} | }
\hline 
\textbf{PS} & \textbf{Research area} & \textbf{Work type} & \textbf{Venue quality} & \textbf{Pub. year} \\ \hline

\citePS{PS11} & Cleaning & Model & \centering NR & 2011  \\ \hline
\citePS{PS57} & Cleaning & Model / Ontology & \centering NR & 2014  \\ \hline
\citePS{PS41} & Cleaning & Model / Ontology & \centering B &  2018  \\ \hline
\citePS{PS38} & Data Quality & DQ Methodology & \centering B &  2015  \\ \hline
\citePS{PS39} & Linked Data & Framework/Model  & \centering B &  2019  \\ \hline
\citePS{PS19} & Data Warehouse & Framework/Model & \centering NR &  2016 \\ \hline
%\citePS{PS40} & Recommender systems & Model & \centering C & 2010 \\ \hline
\end{tabular}
\caption{Main characteristics of the PS that present a context formal definition.}
\label{tab:tableFormalProposal_I}
\end{center}
\end{table}

%% file: _answers.tex
\section{Answers to Research Questions} %work in progress
\label{sec:answers}

%\cvero{Here, you may need a subsection or subsubsection per RQ. Each type of proposal should be described. The idea is not to describe each PS per se, but to describe the part of the PS that is conserned by the question. Then, each PS can be described in several places.}

This section presents the answers brought to the three research questions by the analysis of the selected PS.

\subsection{RQ1: How is context used in data quality models?}

In the PS that address DQ models \citePS{PS2,PS3,PS13,PS18,PS20,PS27,PS28,PS33,PS34,PS59}, we mainly identify that they include context i) distinguishing contextual DQ dimensions from non-contextual ones, ii) defining contextual DQ metrics, which include context components or are delimited by DQ thresholds, and iii) considering the specific context through DQ requirements, which can come from business rules, users' data or user preferences. Next, we will concentrate on analyzing each of the proposals.

To begin we consider the works in \citePS{PS3,PS18}, where are proposed quality-in-use models (3As and 3Cs respectively). They claim that ISO/IEC 25012 DQ model \cite{25012-Stand}, devised for classical environments, is not suitable for Big Data projects, and present \textit{Data Quality in use} models. They propose DQ dimensions as categories called contextual adequacy and contextual consistency, among others, evidencing the need for explicit context consideration. Regarding contextual DQ metrics, in the case of \citePS{PS3}, they also mention that to measure DQ in use in a Big Data project, DQ requirements must be established. In addition, in \citePS{PS18} it is mentioned that DQ dimensions that address DQ requirements of the task at hand should be prioritized.

In the proposal of \citePS{PS2}, the authors reuse the DQ framework of Wang \& Strong \cite{wang-strong} to highlight contextual characteristics of DQ dimensions as completeness, timeliness and relevance, among other. Besides, they underline that contextual DQ increases the retrieval of valuable information from data. Also in \citePS{PS13}, the contextual DQ dimensions included in the proposed DQ model are taken from the bibliography, but in this case the ISO/IEC 25012 standard \cite{25012-Stand} is considered. As well as, the authors claim that DQ requirements play an important role in defining a DQ model, because they depend on the specific context of use. \citePS{PS20} references the proposal in \citePS{PS3} supporting the need raised in it. On the basis that DQ assessment model based-in-use is more and more important, since as in \citePS{PS13}, business value can only be estimated in its context of use. In this case, DQ requirements are strongly tied to the contextual DQ dimensions efficiency and adequacy.

In fact, the proposal in \citePS{PS27} is also motivated by producing value from Big Data analysis minimizing DQ problems. In turn, this work also considers the quality-in-use models in \citePS{PS3,PS18} (3As and 3Cs respectively), but in this case the authors underline that, for these works and others, analyzing DQ only involves preprocessing of Big Data analysis. As well as, they argue that these DQ models mainly consider DQ on a single source, and they do not take sufficiently account user preferences. Hence, the authors present their proposal as a more complete DQ model, because it alerts about DQ problems during the analysis stage in Big Data without any preprocessing, and takes into account user preferences. On the other hand, completeness and consistency are the DQ dimensions considered in this DQ model, and they are contextual according to the users' perspective. Besides, DQ metadata obtained with DQ metrics associated to the DQ dimensions are limited by thresholds specified by users. Therefore, these thresholds can change for each user.

Contrary, results of \citePS{PS28} that presents a DQ profiling model, prove that DQ profiling traces quality at the earlier stage of Big Data life cycle leading to DQ improvement. This proposal profiles a dataset in a DQ domain defined by a set of DQ requirements and data filtering needs. In other matters, motivated by decision making, in \citePS{PS34} along with DQ measurement, DQ problems are also the focus, and they are associated with DQ dimensions. In turn, users DQ requirements give context to the DQ dimensions. It is also added that the lack of combination of general and specific DQ dimensions for analysing DQ affects data fit for uses. As well as, the authors even point out that although data cleaning produces DQ improvement in the short term, it does not have a radical effect on DQ. Therefore, constant DQ improvement is necessary.

To finish, we emphasize that the bibliography supports the fact that the term \textit{quality} depends highly on the context in which it is applied, and to assess DQ for a specific usage, DQ requirements must be described and the compliance of them have to be checked \citePS{PS33,PS59}. Moreover, the same data may be checked for its accordance with reference to different DQ requirements \citePS{PS59}.

\paragraph{Summary} There is vast evidence that DQ assessment is context-dependent. Since several research domains as Linked Data, Decision Making, Big Data and especially DQ domain, present arguments of the importance of having DQ metrics that adapt to the needs of each reality. The bibliography claims that the current DQ models do not take into account such needs, and particular demands of the different application domains, in particular in the case of Big Data. Perhaps a common DQ model is not possible, since each DQ model should be defined taking into account particular characteristics of each application domain. In addition, there is an agreement on the influence of DQ requirements on a contextual DQ model, since according to the literature, they condition all the elements of such model.

\subsection{RQ2: How is context used within quality metrics for the main data quality dimensions?}

Already in subsection \ref{sec:slr:results} we have addressed contextual DQ metrics that we have identified in the analyzed PS. Now, looking for answering this research question, we return to these PS, for a more detailed analysis. Now, we want to know why authors consider certain DQ metrics are contextual, which context components are considered, and how they are included in the definition of DQ metrics. Next, we present this analysis.

The proposals in \citePS{PS16,PS32} are based on the \textit{fitness for use} approach \cite{wang-strong} to the DQ measurement task. \citePS{PS16} defines DQ metric as objective o subjective, the latter when they are based on qualitative evaluations by information and/or data users. It is also mentioned that DQ measurement allows the comparison of the effective quality with predefined DQ requirements. Furthermore, in the case of \citePS{PS32}, the authors underline that DQ requirements have a very important role when implementing a DQ projects, because it should meet the specified DQ requirements. In turn, in that task it is difficult to select appropriate DQ dimensions and their DQ metrics, since there is no agreement on the dimensions that exactly determine DQ. Actually, the previous idea is reinforced by the researchers of \citePS{PS35}, where they point out that DQ requirements may influence over the selection of the set of dimensions and metrics to be considered for DQ assessment. In addition, as a conclusion of a literature review in \citePS{PS21}, the authors define DQ requirements as ``the specification of a set of dimensions or characteristics of DQ that a set of data should meet for a specific task performed by a determined user''. According with this, when we specify DQ characteristics some elements that can define a context must be taken account: such as the specific task, the user, business rules, etc.

Taking Big Data quality issues into account, a proposal of context-dependent DQ assessment in \citePS{PS4} presents a DQ metric for evaluating the confidence precision based on \textit{DQ requirements} specified by users. At the same time, this DQ metric is defined based on certain DQ dimensions such as completeness and distinctness. In a review \citePS{PS17}, the authors recommend an evaluation scheme in which DQ metrics are selected according to DQ dimensions too, beside data and Big Data attributes. On the other hand, \citePS{PS28} proposes a Big Data quality profile repository that includes DQ requirements. This repository defines DQ dimensions and their DQ metrics. Besides, DQ evaluation includes data sources (with all that they imply: data, metadata, etc.), DQ requirements, and DQ evaluation scenarios. Also, but in Data Integration domain, \citePS{PS55} presents users with different roles that specify DQ requirements that later will determine the selection of DQ metrics. 

In the proposals \citePS{PS27,PS39} DQ requirements are represented as threshold. In the former measurement methods use threshold (called quality limit) with which the system alerts users. These thresholds are specific for each measurement method and can be indicated by users. The latter uses thresholds specified by users to condition data sources selection. Data sources whose quality values do not verify the thresholds associated with each DQ dimension are discarded. In the case of \citePS{PS42}, a set of DQ requirements for DQ metrics is specially defined, some of them stating that it must be possible to adjust a DQ metric to a particular \textit{application domain}. In fact, the latter is verified by \citePS{PS15,PS43}, since in \citePS{PS15} the authors define DQ dimensions and DQ metrics whose definition and process of measurement inherently depend on the application domain, resulting in a class of subjective DQ dimensions and DQ metrics. While in \citePS{PS43} it is introduced a set of quality checks for creating application domain specific DQ metrics. According to \citePS{PS38}, a quality criterion might be evaluated by multiple measures, depending on the information characteristics. Then, they evaluate DQ based on the application domain, and the interaction between the user and the application is also considered for contextualizing DQ.

In the case of \citePS{PS43}, \textit{data filtering needs} are included in the definition of DQ metrics, and they are customized by users. In other matters, the proposal in \citePS{PS51} presents DQ metrics that are created using \textit{business rules} that represent conditional functional dependencies. Furthermore, \citePS{PS3} presents a 3As DQ-in-Use model where DQ dimensions (called DQ characteristics by the authors) suggested for Big Data analysis are contextual adequacy, temporal adequacy and operational adequacy. To measure the levels of Data Quality-in-Use, DQ requirements are considered to select the appropriate type of adequacy. Additionally, business rules are used as input to the DQ metrics condition the measurement.

For decision making, a methodology for DQ assessment in \citePS{PS9}, defines DQ metrics based on the \textit{task at hand} (called use-case in this work), data attributes and tests criteria. The latter are used for characterizing datasets and DQ dimensions. With the same purpose the authors of \citePS{PS14} introduce DQ metrics for accuracy of a relational database. The syntactic accuracy assessment matches tuples from the table under evaluation with tuples of another table which contains the same but correct tuples. In this case, DQ metrics are defined based on \textit{other data} that works as a referential. Also taking into account the relational model, the proposals in \citePS{PS19, PS25} are motivated by DQ assessment, but in this case in a Data Warehouse. For the measurement, other data, which are not contextualized data, are taken into account to define the context considered in DQ metrics. In particular, data in the DW dimensions are embedded in the contextual DQ metrics. As well as, in \citePS{PS25}, information from business rules and about the application domain are also be embedded in DQ metrics. 

The approaches in \citePS{PS22,PS31} are different from previous PS, since in these cases the authors investigate how DQ metrics for particular DQ dimensions can be aggregated to derive an indicator for DQ. In the case of \citePS{PS22} values of completeness, validity, and currency are aggregated to derive an indicator for the dimension accuracy. While in \citePS{PS31}, an indicator function is designed as a product of the results of the DQ metrics for completeness, readability and usefulness. One more time, DQ metrics are raised based on other data, which in this case are \textit{DQ metadata}.

\paragraph{Summary}
It is difficult to identify which are the context components that make DQ \textit{fitness for use}. Since literature suggests several context components to contextualize DQ dimensions and DQ metrics. In particular, regarding how context is used within DQ metrics, there is nothing concrete yet. Because in most of the times DQ metrics are considered contextual according to whether the measurements obtained with them conform or not DQ requirements. Actually, we found few works \citePS{PS19,PS22,PS25,PS31, PS43} that explicitly include context components in the definition of DQ metrics.

On the other hand, among all the context components addressed, DQ requirements are the most considered by the different research domains to contextualize DQ. In particular, at the moment to properly select DQ dimensions and define their metrics. This latter is in line with the analysis performed for the RQ1.

\subsection{RQ3: How is context used within data quality concepts?}

To answer this research question and according to our review, we identify DQ concepts that usually consider data context, such as measurement tasks, DQ methodologies, DQ requirements and DQ problems. Next, we will see how the context is related to these DQ concepts.

To begin we consider the proposal in \citePS{PS36}, it takes DQ actions related to business rules. DQ actions, in this case, refer to DQ tasks corresponding to measurement, evaluation and cleaning. Additionally, authors of \citePS{PS11,PS57,PS41} address DQ assessment, focusing on data cleaning and motivated by data filtering needs. In this approach, other data (which are not the evaluated data), DQ requirements and business rules influence DQ evaluations. In other matters, a review carried out in \citePS{PS25}, authors observe that few works use context when performing DQ tasks as data profiling, data cleaning or data evaluation, being DQ measurement one of the tasks that more considers the context. This latter coincides with the results obtained in section \ref{sec:results:ctxComp}, where we observe that it is at the measurement and evaluation stages of a DQ process that the components of the data context are most taken into account. 

On the other hand, in \citePS{PS51} a DQ methodology is proposed for assessing DQ based on business rules. Also \citePS{PS4} proposes a methodology that selects, based on user DQ requirements, the best configuration for DQ assessment. This coincides with arguments of \citePS{PS16}, where is mentioned that the role of DQ methodologies is to guide in the complex decisions to be made, but at the same time, it has to be adapted to the application domain. According to this, \citePS{PS38} considers the application domain and user data for the purpose of presenting reliable information quality. In other matters, we also identify DQ methodologies for particular research domains, for example those proposed by \citePS{PS35, PS36}, which are adapted for assessing DQ in a Big Data Domain. In the case of \citePS{PS25} a methodology is presented to define contextual DQ metrics in Data Warehouse Systems. In other cases, particular methodologies are proposed for DQ assessment. For instance, \citePS{PS23} applies the six-sigma methodology \cite{linderman2003six}, and it addresses DQ tasks (measurement, assessment, and improvement), that are guided especially by DQ requirements. In the case of \citePS{PS13}, a methodology for public sector organizations is based on the OPDCA (Observe, Plan, Do, Check, Adjust/Act) Shewhart-Deming cycle \cite{PDCA}. Besides, DQ requirements, laws and regulations own of the application domain, condition DQ assessment.

Furthermore, \citePS{PS33} presents a theoretical methodology that describes principles of DQ and methods for its evaluation, which are carried out based on DQ requirements. Also in \citePS{PS10,PS56} DQ requirements are the starting point of DQ assessment. In \citePS{PS10} the proposed methodology identifies DQ dimensions and DQ metrics that arise from DQ requirements, focusing on their visualization to assess the overall DQ. While in \citePS{PS56} the methodology identifies DQ problems to perform DQ evaluations, in this case DQ requirements are called specifications. Regarding DQ problems, the authors in \citePS{PS26} highlight that they are an important source to understand data filtering needs. Moreover, they usually result from the violation of DQ requirements. The proposal in \citePS{PS29} asserts that general data problems within a context can result in information quality problems. In particular, the research in \citePS{PS24} classifies DQ problems using a conceptual model to determine an optimal investment in DQ projects. In addition, some of these DQ problems are classified as context dependent. Context is also considered in \citePS{PS9} at the initial stage of a DQ process, where at the final stages DQ is assessed and improved. In this case, the authors emphasize that a specific usage context or data dependent task is defined. As we can see, task and context are used interchangeably.

\paragraph{Summary}

As we have already seen, several works focus on satisfying DQ requirements to achieve DQ tasks as measurement, evaluation and data cleaning. Mainly, DQ requirements vary according to users, applications domains or the task at hand, in particular at the different stages of DQ methodologies. Since each stage proposes different DQ actions. But also, adapted DQ methodologies are suggested for assessing DQ, especially in the Big Data Domain. Additionally, literature suggests that when we analyze general data problems within a context, these could become DQ problems.

%% file: _conclusions.tex
\section{Conclusions}
\label{sec:conclusions}

%\cflavia{To delve into the conclusions and future work}
%\cvero{Definitively. A section about open research questions may emerge.}

This paper reports the protocol and results of a SLR on context in DQ management. The SLR methodology allowed the selection of 58 papers based on their relevance, that were analyzed, with a particular attention on the definition of context in DQ management. In addition, we focused on the main characteristics of the proposals, such as type of work, application domain, considered data model and proposed case study. Furthermore, we identified positive and negative points of each approach, taking them into account for our purpose, which is to provide guidance in the usage of data context in DQ management. 

Regarding the general results of the systematic review, we highlight that the largest number of PS were returned by the digital library Springer, the search string that contains the most important DQ concepts is the one that provided the best results, and the most of the PS are recent, published since 2016. In other matters, models and framework were the types of work most found. In turn, PS that only address DQ domain (called only DQ), were the most common proposals. This makes sense, because our searches specifically pointed to that domain. However, we found several PS of Big Data domain too. Researchers of this community claim that current DQ models do not fit to the needs and characteristics of this research domain. In fact, there are discrepancies among Big Data proposals, since in some cases the authors state that most literature does not address DQ at all stages of a Big Data project, but only at the initial stages. 

With respect to our findings, we note that while the importance of context in DQ management is acknowledged in all the studied PS, only 6 of them present a formal context definition. Among them, 5 correspond to the DQ domain (i.e. only DQ) and 1 corresponds to the Linked Data domain. In turn, half of the selected PS, i.e. 29 PS, do not present any context definition, formal nor informal. In other matters, among the context components identified in our review, we observe that DQ requirements are the most suggested to give context to DQ. To a lesser extent, also data filtering needs, the application domain, business rules, other data (which are not the evaluated data), and the task at hand are suggested to compose the context. 

In relation to DQ processes (we consider a DQ process with 7 stages that is presented in the subsection \label{sec:results:ctxComp}), measurement and evaluation stages seem to be the most contextual, since they are the stages in which more context components are proposed to form the context. With respect to the representation of the context components, rules are the most used. In particular, the rules are written in natural language or using a logical language. Numeric indicators are also used to represent context components, DQ requirements are usually represented as DQ thresholds.

To finish our work, we can conclude, based on the analysis performed in answering our research questions, that there is vast evidence that DQ assessment is context-dependent. Since several research domains argue the importance of having DQ models that suit their needs. This implies that DQ models must be contextual, and therefore DQ dimensions and their DQ metrics. However, in the literature there is no consensus on what exactly makes these DQ concepts contextual. In fact, there is not even agreement on what are the most appropriate DQ dimensions and DQ metrics for a DQ project. The latter makes sense, since due to the contextual nature of DQ, DQ dimensions and DQ metrics must be considered according to the data context of each DQ project. On the other hand, there is a tacit agreement that DQ requirements must be considered in all DQ management tasks, and these can vary according to users, the applications domain and the task at hand.

We also identified that context components are usually proposed for DQ evaluation, where measurements obtained through DQ metrics are compared with quality thresholds that represent DQ requirements. This does not coincide with what was analyzed in the PS, because most of the proposals affirm that the context is necessary both when measuring DQ and when evaluating it. This means that there is awareness of the need of a context, but this is not explicitly addressed, especially when DQ metrics are defined. In fact, although we have identified contextual DQ metrics, they are not defined in a generic way, but for particular domains. Therefore, considering this lack in DQ domain, we are currently working in this direction. In this way, we are modeling the context for DQ management, and at the same time, we are focused on defining a case study that supports the context modeling through definitions of contextual DQ metrics. On the other hand, as future work we are planning to formalize the proposed model and apply it at the different stages of a DQ process.